\documentclass[bib]{ba}

\usepackage{graphicx}
\usepackage{amsfonts,amssymb,amsmath,epsfig,pstricks,caption}
\usepackage[normalem]{ulem}
\usepackage{nicefrac}

\usepackage{bayes}

\begin{document}

\inserttype[ba0001]{article}
\renewcommand{\thefootnote}{\fnsymbol{footnote}}
\author{J. Lee and C. P. Robert}{
 \fnms{Jeong Eun}
 \snm{Lee}
 \footnotemark[1]\ead{jelee@aut.ac.nz}
and
  \fnms{Christian P.}
  \snm{Robert}
  \footnotemark[2]\ead{xian@ceremade.dauphine.fr}
}

\title[ ]{Importance sampling schemes for evidence approximation in mixture models}

\maketitle

\footnotetext[1]{
 Auckland University of Technology, New Zealand
 {\sf jelee@aut.ac.nz}
 
}
\footnotetext[2]{
PSL, Universit\'e Paris-Dauphine, CEREMADE, 
Department of Statistics, University of Warwick, and CREST, Paris
 {\sf  xian@ceremade.dauphine.fr}
}
\renewcommand{\thefootnote}{\arabic{footnote}}

\newcommand\bx{\mathbf{x}}
\newcommand\bz{\mathbf{z}}
\newcommand\bxi{\boldsymbol{\xi}}
\newcommand\blambda{\boldsymbol{\lambda}}

\begin{abstract}

The marginal likelihood is a central tool for drawing Bayesian inference about
the number of components in mixture models. It is often approximated since the
exact form is unavailable. A bias in the approximation may be due to an
incomplete exploration by a simulated Markov chain (e.g., a Gibbs sequence) of
the collection of posterior modes, a phenomenon also known as lack of label
switching, as all possible label permutations must be simulated by a chain in
order to converge and hence overcome the bias. In an importance sampling
approach, imposing label switching to the importance function results an
exponential increase of the computational cost with the number of components.
In this paper, two importance sampling schemes are proposed through choices for
the importance function; a MLE proposal and a Rao--Blackwellised importance
function. The second scheme is called dual importance sampling. We
demonstrate that this dual importance sampling is a valid estimator of the
evidence. To reduce the induced high demand in computation, the original
importance function is approximated but a suitable approximation can produce an
estimate with the same precision and with reduced computational workload. 

\end{abstract}

\section{Introduction}

Consider a sample $\bx=\{x_1,\cdots, x_{n_x}\}$ that is a realisation of a random sample (univariate or
multivariate) from a finite mixture of $k$ distributions
\[
X_j \sim f_k(x|\theta) = \sum_{i=1}^k \lambda_i f(x|\xi_i)\,, \hspace{1cm}  j=1,\cdots , n_x
\]
where the component weights $\blambda=\{\lambda_i\}_{i=1}^k$ are non-negative and sum to $1$. The
collection of the component-specific parameters is denoted $\bxi=\{\xi_i\}^k_{i=1}$ and the collection of all parameters 
by $\theta=\{\blambda,\bxi\}$. As is now standard \citep{marin:mengersen:robert:2007} each observation $x_j$ from the
sample can be assumed to originate from a specific if unobserved component of $f_k$, denoted $z_i$, and
the mixture inference problem can then be analysed as a missing data model, with discrete missing data
$\bz=\{z_1,\ldots,z_{n_x}\}$, such that
\[
x_j|\bz\sim f(x_j|\xi_{z_j})\, , \hspace{1cm} \mbox{independently for } j=1,\cdots, n_x \, .
\]
The conditional distribution of $z_j\in [1,\ldots,k]$ is then given by
\[
z_j|\bx,\theta \sim \mathcal{M}\left( \dfrac{\lambda_1f(x_j|\xi_1)}{\sum_{i=1}^k \lambda_if(x_j|\xi_i)}, \ldots, 
\dfrac{\lambda_k f(x_j|\xi_k)}{\sum_{i=1}^k \lambda_if(x_j|\xi_i)}  \right) \,.
\]

\noindent This interpretation of the mixture model leads to a natural clustering of the observations
according to their label and the cluster associated with the mixture component $i$ provides information about
$\lambda_i$ and $\xi$. In particular, when the full conditional distribution of the parameter $\theta$ is available in
closed form, conditional simulation from $\pi(\bxi,\blambda |\bx,\bz)$ becomes straightforward (see
\citet{diebolt:robert:1994}).

\noindent In a Bayesian mixture modelling setup, the goal is to perform inference on the parameter
$\theta$ and the posterior distribution $\pi(\theta|\bx)$ is usually approximated via MCMC methods. The likelihood
function $p_k(\bx|\theta)$ is both available and invariant under permutations of the component indices. If an
exchangeable prior is chosen on $(\blambda,\bxi)$, the posterior density reproduces the likelihood invariance and
component labels are not identifiable. This phenomenon is called {\em label switching} and is
well-studied in the literature \citep{celeux:hurn:robert:2000,stephens:2000,jasra:holmes:stephens:2005}.
From a simulation perspective, label switching induces multimodality in the target and while it is
desirable that a simulated Markov chain targeting the posterior explores all of the $k!$ symmetric modes of the posterior
distribution, most samplers fail to switch between modes \citep{celeux:hurn:robert:2000}. For instance, when using a
data augmentation scheme, which is a form of Gibbs sampler adapted to missing data problems \citep{robert:casella:2004}, the Markov chain very slowly if ever switches between the symmetric modes. Therefore, since the chain only explores a certain region of the support of the multimodal posterior, estimates based on the simulation output are necessarily biased. When label switching is missing from the MCMC output, it can be simulated by modifying the MCMC sample (see \citet{fruhwirth:2001, papastamoulis:roberts:2008, papastamoulis:illiopoulos:2010}). 

\noindent A different perspective on the label switching phenomenon is inferential. Indeed, point
estimates of the component-wise parameters are harder to produce when the Markov chain moves freely between modes. To
achieve component-specific inference and give a meaning to each component, relabelling methods have
been proposed in the literature (see \citet{richardson:green:1997,celeux:hurn:robert:2000,
stephens:2000,jasra:holmes:stephens:2005, marin:robert:2007, geweke:2007, rodriguez:walker:2014} and others). An
R-package, {\ttfamily label.switching} \citep{papastamoulis:2013}, incorporates some of those label switching removing
methods. 

\noindent Evaluating the number of components $k$ is a special case of model comparison, which can be conducted by
comparing the {\em posterior probabilities of the models}. Those probabilities are in turn computed via the marginal
likelihoods $\mathfrak{E}(k)$, also known as model evidences \citep{richardson:green:1997}
\[ 
\mathfrak{E}(k) = \int_{S} p_k(\bx|\theta) \pi_k(\theta) \, \text{d} \theta ~, 
\]
where $\pi_k(\theta)$ is the selected prior for the $k$-component mixture. (We assume here that it is exchangeable wrt
its components.) Recall that the ratio of evidences is a Bayes factor and is properly scaled to be readily compared to
$1$ \citep{jeffreys:1939}. When a large collection of values of $k$ is considered for model comparison, 
sophisticated MCMC methods have been developed to bypass computing evidences
\citep{richardson:green:1997,stephens:2000a}, even though those are estimated as a byproduct of the methods.
Alternatively, estimating the number of components can also proceed from a Bayesian nonparametric (BNP) modelling, which
assumes an infinite number of components and then evaluates the non-empty components implicitly through partitioning
data, using for instance the Chinese restaurant process \citep{antoniak:1974, escobar:west:1995, rawmussen:2000}. This
however requires a modification of the prior modelling and we will not cover it in this paper, which reassesses Monte
Carlo ways of approximating the evidence.

\noindent The difficulty with approaches using $\mathfrak{E}(k)$ is that the quantity often cannot directly and
reliably be derived from simulations from the posterior distribution $\pi(\theta | \bx)$ \citep{newton:raftery:1994}.
The quantity has been approximated using dedicated methods such harmonic means \citep{satagopan:newton:ratfery:2000,
ratfery:etal:2006}, importance sampling \citep{rubin:1987,rubin:1988,gelman:meng:1998}, bridge sampling
\citep{meng:wong:1996, meng:schilling:2002}, Laplace approximation \citep{tierney:kadane:1986,diciddio:etal:1997},
stochastic substitution \citep{gelfand:smith:1990, chib:1995, chib:1996}, nested sampling \citep{chopin:robert:2010},
Savage-Dickey representations
\citep{verdinelli:wasserman:1995,marin:robert:2010b} and erroneous implementations of the Carlin and Chib algorithm
\citep{carlin:chib:1995,scott:2002,congdon:2006,robert:marin:2008}. Comparative studies of those methods are found in
\cite{marin:robert:2010} and \citet{ardia:etal:2012}. 

\noindent In the specific case of mixtures, the invariance of the posterior density under an arbitrary relabelling of
the mixture components must be exhibited by simulations and approximations to achieve a valid estimate of
$\mathfrak{E}(k)$ as discussed in \citet{neal:1999, berkhof:mechelen:gelman:2003, marin:robert:2008}. This often leads
to computationally intensive steps in approximation methods, especially when $k$ is large, and it is the purpose of this
paper to provide a partial answer to this specific issue.

\noindent We consider here two estimators of $\mathfrak{E}(k)$, both based on importance sampling (IS). One is
a version of Chib's estimator and the second one a novel representation called {\em dual importance sampling}. Our
importance construction aims to better approximate the posterior distribution both around a specific local mode and at
the corresponding $(k!-1)$ symmetric modes of the posterior distribution. A particular mode is first approximated based
on (i) a point estimate and (ii) Rao--Blackwellisation from a Gibbs sequence. Then, the corresponding local density
approximate is permuted to reach all modes. We demonstrate here that dual importance sampling is comparable to our
benchmark method, Chib's approach. Taking advantage of the symmetry in the posterior distribution, we show how to reduce
computational demands by approximating our importance function.

\noindent The paper starts with recalling the approximation techniques of Chib's method and bridge sampling in Section
2. In Section 3, importance sampling is considered, including our choices of importance functions.  Our importance
function approximate approach is introduced in Section 4. Simulation studies using both simulated and benchmark
datasets, namely the galaxy and fishery datasets used in \cite{richardson:green:1997} are reported in Section 5, and the
paper concludes with a short discussion in Section 6. 

\section{Standard evidence estimators}

\subsection{Chib's estimator and corrections}

In this paper, the reference estimator of evidence is Chib's(\citeyear{chib:1995}) method and is derived from rewriting
Bayes' theorem
\begin{equation}\label{eq:chib}
\widehat{\mathfrak{E}}(k) = m_k(\bx) = \dfrac{\pi_k(\theta^o) p_k(\bx|\theta^o)}{\pi_k(\theta^o|\bx)} 
\end{equation}
where $\theta^o$ is any plug-in value for $\theta$. When $\pi_k(\theta^o|\bx)$ is not available in
closed form, the Gibbs sampling decomposition allows a Rao--Blackwellised approximation \citep{gelfand:smith:1990,robert:casella:2004} 
\[
\widehat\pi_k(\theta^o|\bx) = \frac{1}{T} \sum_{t=1}^T \pi_k(\theta^o|\bx,\bz^t)\, , 
\]
where $( \bz^t)_{t=1}^T$ is a Markov chain with stationary distribution $\pi_k(\bz|\bx)$. The appeal of this
estimator, when available, is that it constitutes a non-parametric density estimator converging at a regular parametric
rate.

\noindent It is now an accepted fact that label switching is necessary for the above Rao--Blackwellised
$\hat\pi_k(\theta^o|\bx)$ to converge. When $(\bz^1,\cdots,\bz^T)$ only explores part of the modes of
the posterior, this estimator is biased, generally missing the target quantity $\log(m_k(\bx))$ by a factor of order
$\text{O}(\log\,k!)$, with no simple correction factor \citep{neal:1999}. \cite{berkhof:mechelen:gelman:2003} later
suggested a generic correction by averaging $\widehat\pi_k(\theta^o|\bx)$ over all possible permutations of the labels,
hence forcing ``perfect" label switching. The resulting approximation is expressed as
\[
\tilde\pi_k(\theta^o|\bx) = \frac{1}{T k!}
\sum_{\sigma\in\mathfrak{S}_k}\sum_{t=1}^T \pi_k(\theta^o|\bx,\sigma(\bz^t))\,,
\]
where $\mathfrak{S}_k$ denotes the set of the $k!$ permutations of $\{1,\ldots,k\}$ and $\sigma$ is one of those
permutations. Note that the above correction can also be rewritten as
\begin{equation} \label{eq_01}
\tilde\pi_k(\theta^o|\bx) = \frac{1}{T k!}
\sum_{\sigma\in\mathfrak{S}_k}\sum_{t=1}^T \pi_k(\sigma(\theta^o)|\bx,\bz^t)\,,
\end{equation}
using a notational shortcut $\sigma(\theta^o)$ meaning that the components of $\theta^o$ are then switched according to the permutation $\sigma$. This representation may induce computational gains since only $k!$ versions of $\sigma(\theta^o)$ need to be stored.

While Chib's representation has often been advocated as a highly stable solution for computing the evidence in mixture
models, which is why we selected it as our reference, alternative solutions abound within the literature, including
nested sampling \citep{skilling:2007,chopin:robert:2010}, reversible jump MCMC \citep{green:1995,richardson:green:1997},
and particle filtering \citep{chopin:2002}.

\subsection{Bridge Sampling}

\citet{meng:wong:1996} proposed a sample--based method to compute a ratio of normalizing constants of
two densities with common support. The method is well-suited to estimate the marginal likelihood
\citep{fruhwirth:2001,fruhwirth:2004} and used as a point posterior estimate for Chib's method
\citep{mira:nicholls:2004}. Considering a normalised density $q$ and the unnormalized posterior distribution
$\pi_k^*(\theta|\bx)=\pi_k(\theta)p_k(\bx|\theta)$, the bridge sampling identity is given by
\[
\widehat{\mathfrak{E}}(k) =
\dfrac{\mathbb{E}_{q(\theta)}[\alpha(\theta)\pi_k^*(\theta|\bx)]}{\mathbb{E}_{\pi_k(\theta|\bx)}[\alpha(\theta)q(\theta)]}
\,,
\]
for an arbitrary function $\alpha$ (provided all expectations are well-defined, \citealp{chen:shao:ibrahim:2000}). The
(formally) optimal choice for $\alpha$ \citep{meng:wong:1996} leads to the following iterative estimator
\begin{equation} \label{eq_BS}
\widehat{\mathfrak{E}}^{(t)}(k) = \widehat{\mathfrak{E}}^{(t-1)}(k)\, 
\dfrac{M_1^{-1}\displaystyle\sum^{M_1}_{l=1}\nicefrac{\hat{\pi}_{t-1}(\tilde{\theta}^{l}|\bx)}
{M_1 q(\tilde{\theta}^{l})+M_2\hat{\pi}_{t-1}(\tilde{\theta}^{l}|\bx)}}
{M_2^{-1}\displaystyle\sum^{M_2}_{m=1}\nicefrac{q(\hat{\theta}^{m})}{M_1 q(\hat{\theta}^{m})+M_2\hat{\pi}_{t-1}(\hat{\theta}^{m}|\bx)}} 
\end{equation}
where $\hat{\pi}_{t-1}(\theta|\bx)=\pi_k^*(\theta|\bx)/\widehat{\mathfrak{E}}^{(t-1)}(k)$. Here,
$(\tilde{\theta}^1,\ldots,\tilde{\theta}^{M_1})$ and $(\hat{\theta}^1,\ldots,\hat{\theta}^{M_2})$ are samples from
$q(\theta)$ and $\pi_k(\theta|\bx)$ respectively.

\noindent The convergence of bridge sampling (with the above optimal $\alpha$) is trivial when
$\pi_k^*(\theta|\bx)$ and $q(\theta)$ share a sufficiently large portion of their supports. If the
support intersection is too small, the resulting bridge sampling estimator may end up with an infinite variance
\citep{voter:1985, servidea:2002}. Improvements of the algorithm, like path sampling \citep{gelman:meng:1998}, a simple
location shift of the proposal distribution \citep{voter:1985}, and a warp bridge sampling \citep{meng:schilling:2002},
have been proposed. 

\noindent In the specific case of the mixture posterior distribution, the parameter $\theta$ can be split in
$\blambda$ and $k$ further blocks $\xi_1,\ldots,\xi_k$ in the Gibbs sampling steps. The output samples from the Gibbs
sampler are denoted by $\{\theta^{(j)},\bz^{(j)}\}^{J_1}_{j=1}$, where the $\bz^{(j)}$'s are the component allocation vectors
associated with the observations $\bx$. For bridge sampling, \citet{fruhwirth:2004} suggested using a Rao--Blackwellised function
$q(\theta)=q(\blambda,\bxi)$ of the form 
\begin{eqnarray}
q(\theta) &=&  \frac{1}{J_1}\sum_{j=1}^{J_1} \pi_k(\theta| \theta^{(j)}, \bz^{(j)},\bx) \label{eq_perms_q} \\
&=& \frac{1}{J_1}\sum_{j=1}^{J_1} p(\blambda| \bz^{(j)}) \prod^k_{i=1} p(\xi_i|\bxi^{(j)}, \bz^{(j)},\bx)  \nonumber 
\end{eqnarray} 
assuming $\{ \theta^{(j)}, \bz^{(j)} \}_{j=1}^{J_1}$ is well-mixed, followed by switching the labels of the simulations
from the posterior distribution \citep{fruhwirth:2001}.  \citet{fruhwirth:2004} demonstrated that the iterative bridge
sampling estimator (\ref{eq_BS}), using (\ref{eq_perms_q}) as $q(\cdot)$, converges relatively quickly, in about $t=10$
iterations, even with different starting values.  

\section{New importance sampling estimators}

If $q(\theta)$ is an importance function with support $S_q$, generating a sample $\boldsymbol{\theta}=(\theta^{(1)},\ldots,\theta^{(T)})$ from $q(\theta)$ leads to the evidence approximation 
\begin{equation}
\widehat{\mathfrak{E}}(k) = \frac{1}{T} \sum^T_{t=1} \dfrac{\pi_k(\theta^{(t)})p_k(\bx|\theta^{(t)})}{q(\theta^{(t)})}
\stackrel{\text{def}}{=} \frac{1}{T} \sum^T_{t=1} \omega(\theta^{(t)}) \,. \label{ImpSamp}
\end{equation}
To provide a good approximation, the support of $q(\theta)$ must overlap the support of the posterior distribution,
which is both symmetric under permutations and multimodal. In this sense, a Rao--Blackwellised estimate like
(\ref{eq_perms_q}) is a natural solution for the choice of $q$, despite the drawback that $J_1$ increases ``factorially"
fast with $k$ due to the permutations over $\{\theta^{(j)},\bz^{(j)}\}_{j=1}^{J_1}$ \citep{fruhwirth:2004,
fruhwirth:2006}. 

\noindent In the following sections, the parameter $\theta$ is decomposed into $(k+1)$ blocks
$\theta=(\blambda,\xi_1,\ldots,\xi_k)$. Note that $\xi_i$ is a component-wise block, most often a vector.
Two types of importance functions, based on the product of marginal posterior distributions, will be
considered. The usefulness and details of the product of block marginal posterior distributions are well summarised in
\citet{perrakie:etal:2014}.

\subsection{A plug-in proposal}

Using a very simple Rao--Blackwell argument inspired from Chib's representation, a natural importance function is 
$$
q(\theta) = \pi_k(\theta|\bz^o,\theta^o,\bx).
$$
Samples are generated from the posterior distribution conditional on a given completion
vector $\bz^o$, which is usually taken equal the MAP (maximum a posteriori) or the marginal MAP estimate of $\bz$
derived from MCMC simulations. Taking the full permutation of component labels of $\bz^o$ and
$\theta^o$ (inspired by \cite{berkhof:mechelen:gelman:2003} and \cite{marin:robert:2008}), we thus propose a symmetrised version of a MAP proposal
\begin{eqnarray}
q(\theta) &=& \frac{1}{k!} \sum_{\sigma\in\mathfrak{S}_k} \pi_k(\theta| \sigma(\theta^o,\bz^o),\bx) \label{eq_q_mle} \\
&=& \frac{1}{k!} \sum_{\sigma\in\mathfrak{S}_k} p(\blambda| \sigma(\bz^o)) \prod^k_{i=1} p(\xi_i|\sigma(\bxi^o), \sigma(\bz^{o}),\bx)  \,. \nonumber  
\end{eqnarray}
This proposal is equivalent to generating $\theta$ from $\pi_k(\theta|\theta^o,\bz^o,\bx)$ and then operating a random
permutation on the components of $\theta$. The computational cost of producing $\omega(\theta)$ in \eqref{ImpSamp}, hence
$\widehat{\mathfrak{E}}(k)$, is then multiplied by $k!$ under the provision that the support of (\ref{eq_q_mle}) is
sufficiently wide. If the tails of samples generated from (\ref{eq_q_mle}) are deemed to be too narrow, as signalled by
the effective sample size, additional selected (and thinned) simulations $\bz^1,\ldots,\bz^t$ taken from the Gibbs
output can be included to make the proposal more robust.

\noindent While this estimator is theoretically valid, providing an unbiased estimator of $\widehat{\mathfrak{E}}(k)$, it may face difficulties in practice by missing wide regions of the parameter space when simulating from $\pi_k(\theta|x,z^o)$. This is indeed the practical version of simulating from an importance function with a support that is smaller than the support of the integrand a setting that leads to an erroneous approximation of the corresponding integral. In the
current situation, since $\pi_k(\theta|x,z^o)$ is everywhere positive, this is not a theoretical issue. However, in
practice, the conditional density is numerically equal to zero around the alternative modes. \\

\subsection{Dual importance sampling}

A dual exploitation of the above Rao--Blackwellisation argument produces an alternative importance sampling proposal, based on MCMC draws $\{\theta^{(j)}, \bz^{(j)} \}^J_{j=1}$ from the unconstrained posterior
distribution. 
The new importance function is expressed as
\begin{eqnarray}
q(\theta) &=& \frac{1}{J k!}\sum_{j=1}^J \sum_{\sigma\in\mathfrak{S}_k} \pi_k( \theta| 
\sigma( \theta^{(j)} , \bz^{(j)} ),\bx)  \label{eq_DS} \\
&=& \frac{1}{J k!}\sum_{j=1}^J \sum_{\sigma\in\mathfrak{S}_k } p(\blambda| \sigma(\bz^{(j)})) \prod^k_{i=1} p(\xi_i|\sigma(\bxi^{(j)}) , \sigma(\bz^{(j)}),\bx)  \,. \nonumber  
\end{eqnarray}

\noindent Here, $\pi_k( \theta| \sigma(\theta^{(j)},\bz^{(j)}),\bx)$ is a product of full conditional densities on each 
parameter in a Gibbs sampler representation and $\{\theta^{(j)},\bz^{(j)} \}^J_{j=1}$ is the original albeit not
necessarily well-mixed simulation outcome. Label switching is imposed upon those $J$ conditional densities through all
$k!$ permutations and conversely the average of $J$ well-selected conditional densities should approximate the posterior
around any of the $k!$ symmetric modes of this posterior.

\noindent If we now assume that the component labels of the terms $\{\theta^{(j)},\bz^{(j)} \}_{j=1}^J$ in
(\ref{eq_DS}) have not been permuted and that any label switching occurence has been removed from the simulations by a
recentering method \citep{celeux:hurn:robert:2000}, we denote the resulting transforms by $\{\varphi^{(j)}\}_{j=1}^J$. They can be
interpreted as hyperparameters of $q$. The density (\ref{eq_DS}) then satisfies
\begin{equation}
q(\theta) = \dfrac{1}{Jk!} \sum_{j=1}^J \displaystyle\sum_{i=1}^{k!}  \pi(\theta |\sigma_i(\varphi^{(j)}),\bx) 
\stackrel{\text{ef}}{=} \dfrac{1}{k!} \displaystyle\sum_{i=1}^{k!} h_{\sigma_i}(\theta)  
\label{eq_DS_1}
\end{equation}
where $h_{\sigma_i}(\theta) = \dfrac{1}{J} \displaystyle\sum_{j=1}^J \pi(\theta |\sigma_i(\varphi^{(j)}),\bx)$. Each of
the densities $h_{\sigma_1},\cdots, h_{\sigma_{k!}}$ has a support--i.e., a domain where it takes non-negligible
values-- denoted by $S_{\sigma_1},\cdots, S_{\sigma_{k!}}$ and
$S_q = \bigcup_{i=1}^{k!} S_{\sigma_i}$. Note that an estimator using (\ref{eq_DS_1}) is equivalent to an estimator
using (\ref{eq_DS}).

\noindent From a computational perspective, an artificial label switching step is necessary in computing $q(\theta)$ but
not in generating a proposal $\theta$ from $q$. For arbitrary permutation representations $\sigma_m,\sigma_c,\sigma_i\in
\mathfrak{S}_k=\{\sigma_1,\dots, \sigma_{k!}\}$ acting on both $\theta$ and $\varphi$, the following holds for (\ref{eq_DS})
$$ 
\pi(\sigma_c(\theta)|\sigma_i(\varphi),\bx) = \pi(\sigma_m\sigma_c(\theta)|\sigma_m
\sigma_i(\varphi),\bx) \,, 
$$
where $\sigma_m\sigma_c(\theta)=\sigma_m(\sigma_c(\theta))$. The full permutation representation set is invariant over
an additional permutation representation $\sigma_m$ (e.g., $\mathfrak{S}_k=\{\sigma_m\sigma_1,
\cdots,\sigma_m\sigma_{k!}\}$), $q(\sigma_c(\theta))$ and $q(\sigma_m\sigma_c(\theta))$ are equal. Thus the standard
estimator using $q$ in (\ref{eq_DS}) is equivalent (from a computational viewpoint) to an estimator such that (i)
proposals are generated from a particular term $h_{\sigma_c}(\theta)$ of (\ref{eq_DS_1}) and (ii) importance weights are
computed according to (\ref{eq_DS_1}). This makes a proposal generating step easier by ignoring label switching even
though all the $h_{\sigma}(\theta)$'s need be evaluated to compute $q(\theta)$.

\subsection{Importance function based on marginal posterior densities}

\noindent Importance functions found in (\ref{eq_perms_q}) and (\ref{eq_DS_1}) have the same underlying motivation of a
better approximation of the joint posterior distribution and the resulting estimate of (\ref{ImpSamp}) should therefore be
more efficient. Both are designed to cover the $k!$ clusters, which are created by either
symmetrizing the labels of hyperparameter set $\{ \theta^{(j)},\bz^{(j)}\}_{j=1}^{J}$ as in (\ref{eq_DS_1}) or by
randomly permuting the label of each $\{\theta^{(j)},\bz^{(j)}\}_{j=1}^{J_1}$ as in (\ref{eq_perms_q}). Once $k!$
clusters of hyperparameters are thus constructed, the corresponding conditional densities constitute clusters for $q$.  

\noindent Consider $\kappa\in \{1,\ldots,k!\}$, a cluster index of $q$. Associating the cluster component function
$q_{\kappa}(\cdot|\bx)$ with a support $S_{\kappa}$, the importance function $q$ is expressed as 
\begin{equation} \label{eq_q_cluster} 
q(\theta|\bx) = \sum^{k!}_{\kappa=1} p(\kappa) q_{\kappa}(\theta|\bx)
\end{equation}
where $p(\kappa)$ is the proportion of those conditional densities that are associated with the cluster $\kappa$ and
$\sum_{\kappa=1}^{k!} p(\kappa)=1$. The importance function representation (\ref{eq_DS_1}) is thus a special case of
\eqref{eq_q_cluster} with ($\kappa=1,\ldots,k!$)
$$
S_{\sigma_{\kappa}}=S_{\kappa}\,,\ h_{\sigma_{\kappa}}(\theta)=q_{\kappa}(\theta|\bx)\ \text{ and }\ p(\kappa) = 1/k!\,. 
$$
By contrast, the density (\ref{eq_perms_q}) does not achieve perfect symmetry, which means
$\kappa$ is not uniformly distributed, although $p(\kappa) \to 1/k!$ as $J_1\to \infty$.

\noindent A marginal likelihood estimate using $q(\theta)$ as in (\ref{eq_q_cluster}) follows by a
standard importance sampling identity
 
\begin{eqnarray}
\mathfrak{E}(k) &=& \displaystyle\int_{S_q} \dfrac{\pi(\theta)
p_k(\bx|\theta)}{q(\theta|\bx)}  \left(\sum_{\kappa=1}^{k!} p(\kappa)q_{\kappa}(\theta|\bx) \right) \text{d} \theta \nonumber \\
&=&  \displaystyle\sum_{\kappa=1}^{k!} \int_{S_{\kappa}} \dfrac{\pi(\theta)
p_k(\bx|\theta)}{q(\theta|\bx)} p(\kappa)  q_{\kappa}(\theta|\bx) \text{d} \theta = \mathbb{E}_{p(\theta,\kappa)}[\omega(\theta)]  \label{eq_IS1}
\end{eqnarray}
leading to
\[
\widehat{\mathfrak{E}}(k) = \dfrac{1}{T}\sum_{t=1}^T \omega(\theta^{(t)})\,, 
\]
where $\omega(\theta)={\pi(\theta) p_k(\bx|\theta)}\big/{q(\theta|\bx)}$, namely
a weighted sum of integrals over the $S_{\kappa}$'s ($\kappa=1,\ldots,k!$).

\noindent Due to the perfect symmetry in the clusters of (\ref{eq_DS_1}), the integrals of $\omega q_{\kappa}$ with respect to $\theta$ over $S_{\kappa}$ for $\kappa=1,\cdots,k!$ are equal. Given an arbitrary cluster, $\kappa^o$, the evidence is 
\begin{eqnarray}
\mathfrak{E}(k) &=& \displaystyle\sum_{\kappa=1}^{k!} p(\kappa)  \left(\displaystyle\int_{S_{\kappa}} \omega(\theta)  q_{\kappa}(\theta|\bx) \text{d} \theta \right)  \nonumber \\
&=& \displaystyle\int_{S_{\kappa^o}} \omega(\theta) q_{\kappa^o}(\theta|\bx) \text{d} \theta = \mathbb{E}_{q_{\kappa^o}(\theta|\bx)} [\omega(\theta)]  ~. \label{eq_IS2} 
\end{eqnarray}
Note that the corresponding estimator (Monte Carlo approximation based on $T$ draws) for the above is exactly in the same form to the estimator for (\ref{eq_IS1}). 

\noindent Both (\ref{eq_IS1}) and (\ref{eq_IS2}) are thus importance sampling estimators using
(\ref{eq_perms_q}) and (\ref{eq_DS_1}) respectively. Hence standard convergence result hold: by the Law of Large
Numbers, both estimates a.s.~converge to $\mathfrak{E}(k)$, and the Central Limit theorem also holds  
\[ 
\sqrt{T} \left\{  \frac{1}{T}\sum_{t=1}^T \omega(\theta^{(t)}) - \mathfrak{E}(k) 
\right\}  \mathrel{\mathop{\longrightarrow}_{T\to \infty}} \mathcal{N} (0,V_1)
\,,\ \ 
\sqrt{T}\left\{ \frac{1}{T} \sum_{t=1}^T \omega(\theta^{(t)}) - \mathfrak{E}(k) 
\right\} \mathrel{\mathop{\longrightarrow}_{T\to \infty}} \mathcal{N} (0,V_2)
\]
where $V_1=\text{var}_{p(\theta,\kappa|\bx)}(\omega(\theta))$ and
$V_2=\text{var}_{q_{\kappa^o}(\theta|\bx)}(\omega(\theta))$. The perfect symmetry in the clusters of \eqref{eq_DS_1}
does not guarantee a better efficiency in estimation and those variances are rather highly related to how well the
importance functions approximate the joint posterior distribution. If $J_1=Jk!$ and both importance functions provide a
good approximation, $V_1 \approx V_2$ is expected.

\section{Importance function approximation}

Both estimators (\ref{eq_IS1}) and (\ref{eq_IS2}) suffer from massive computational demands when $k$ is large. In this
section, we show how to approximate (\ref{eq_DS}) and increase the computational efficiency (i.e., computing time) as a result.

\noindent It was shown in Section 3.2 that $q$ as in (\ref{eq_DS}) is invariant under a permutation of the labels of
$\theta$ and that proposals can be generated from a particular term $h_{\sigma_c}(\theta)$ of (\ref{eq_DS_1}) without any
loss of statistical efficiency. Given $(\theta^{(1)},\ldots,\theta^{(T)}) \sim h_{\sigma_c}(\theta)$, it is
natural to consider whether or not all terms in $\{h_{\sigma_1}(\theta^{(t)}),\dots,h_{\sigma_{k!}}(\theta^{(t)})\}$ are
different from zero for $t=1,\ldots,T$. In the case some are not, it is obviously computationally relevant to determine
which ones among them are likely to be insignificant (i.e., almost zero). This perspective motivates the following section.

\subsection{Dual importance sampling using an approximation}

Given a proposal $\theta$ generated from a particular 
$h_{\sigma_c}(\theta)$, $\theta\in S_{\sigma_c}$, the importance function at $\theta$ is an average of all
$h_\sigma(\theta)$'s and the relative contribution of each term is
$$ 
\eta_{\sigma_i}(\theta) = {h_{\sigma_i}(\theta)}\big/{k! q(\theta)} 
= {h_{\sigma_i}(\theta)}\Big/{\sum_{l=1}^{k!} h_{\sigma_l}(\theta)} ~, \hspace{1cm} 
i=1,\dots,k! ~.
$$ 
If $\eta_{\sigma_i}(\theta)$ is close to zero, $h_{\sigma_i}(\theta)$ is negligible within $q(\theta)$ and on the opposite
$\eta_{\sigma_i}(\theta) \approx 1$ indicates a high contribution of $h_{\sigma_i}(\theta)$. The expected relative
contribution of $h_{\sigma_i}(\theta)$ 
$$ 
\mathbb{E}_{h_{\sigma_c}}[\eta_{\sigma_i}(\theta)] = \displaystyle\int_{S_{\sigma_c}} 
\eta_{\sigma_i}(\theta) h_{\sigma_c}(\theta)\, \text{d}\theta \label{eq_E_eta}
$$ 
is estimated by
\begin{equation}
\widehat{\mathbb{E}}_{h_{\sigma_c}}[\eta_{\sigma_i}(\theta)] = 
\dfrac{1}{M}\displaystyle\sum^M_{l=1} \eta_{\sigma_i}(\theta^{(l)}) 
~, \hspace{1cm} \theta^{(l)}\sim h_{\sigma_c}(\theta) ~. \label{eq_eta}
\end{equation}
After an appropriate permutation of the indices, we obtain that
$\widehat{\mathbb{E}}_{h_{\sigma_c}}[\eta_{\sigma_1}(\theta)]\geq \cdots \geq
\widehat{\mathbb{E}}_{h_{\sigma_c}}[\eta_{\sigma_{k!}}(\theta)]$, namely that the corresponding $h_{\sigma_1},\cdots,
h_{\sigma_{k!}}$ are in decreasing order of expected contributions. The importance function $q(\theta)$ can then be approximated
by using only the $n$ most important $h_\sigma$'s ($1\leq n \leq k!$), leading to the approximation
\begin{equation} \label{eq_08}
\tilde{q}_n(\theta) = \dfrac{1}{k!}\displaystyle\sum^n_{i=1} h_{\sigma_i}(\theta)~,
\end{equation}
and the mean absolute difference from $q(\theta)$ is approximated by
\begin{equation} \label{eq_AbsErr}
\hat{\phi}_n = \dfrac{1}{M}\displaystyle\sum_{l=1}^M \left |\tilde{q}_n(\theta^{(l)}) - q(\theta^{(l)}) \right | ~, \hspace{1cm} \theta^{(l)}\sim h_{\sigma_c}(\theta) ~.
\end{equation}

\noindent When this mean absolute difference is below a certain threshold, $\tau$, $\tilde{q}_n$ is considered to be an appropriate approximation for $q$. We define the corresponding approximate set $\mathfrak{A}(k) \subseteq\mathfrak{S}_k$ to be made of $\{\sigma_1,\cdots, \sigma_n\}$, $n$ being defined as the smallest size that satisfies the condition $\widehat{\phi}_n<\tau$. With this truncation, the computational efficiency obviously improves.

\noindent Note that the set $\mathfrak{A}(k)$ is determined under the assumption that all proposals
($\theta^{(t)}$) are generated from $h_{\sigma_c}$ since the quality of the approximation is only
guaranteed under this assumption. Due to the perfect symmetry of $q(\theta)$ over
the $k!$ permutations, the choice of $\sigma_c$ is obviously irrelevant for the computational gains.
The evidence estimate using such an approximation is detailed in the following algorithm:\\

\fbox{
  \parbox{\textwidth}{
    {\bf Algorithm 1: Dual importance sampling algorithm with approximation}
    
\begin{description}
\item[1] Randomly select $\{\bz^{(j)}, \theta^{(j)} \}^J_{j=1}$ from Gibbs sample and remove label switching by an appropriate method. 
Then, construct $q(\theta)$ as in (\ref{eq_DS_1}). 

\item[2] Derive the corresponding term $h_{\sigma_c}(\theta)$ and generate particles $\{\theta^{(t)}\}^T_{t=1} \sim h_{\sigma_c}(\theta)$.

\item[3] Construct an approximation, $\tilde{q}(\theta)$, using the first $M$ terms in $\{\theta^{(t)}\}^T_{t=1}$:
\begin{description}
\item[3.1] Compute $(h_{\sigma_1}(\theta^{(t)}),\dots,
h_{\sigma_{k!}}(\theta^{(t)}),\eta_{\sigma_1}(\theta^{(t)}),\dots, \eta_{\sigma_{k!}}(\theta^{(t)}))$ for $t=1,\dots,M$
and $\widehat{\mathbb{E}}_{h_{\sigma_c}}[\eta_{\sigma_1}(\theta)], \cdots,
\widehat{\mathbb{E}}_{h_{\sigma_c}}[\eta_{\sigma_{k!}}(\theta)]$ as in (\ref{eq_eta}).

\item[3.2] Reorder the $\sigma$'s so that 
$\widehat{\mathbb{E}}_{h_{\sigma_c}}[\eta_{\sigma_1}(\theta)]\geq \cdots \geq
\widehat{\mathbb{E}}_{h_{\sigma_c}}[\eta_{\sigma_{k!}}(\theta)]$.

\item[3.3] Initialise $n=1$ and compute $\tilde{q}_n(\theta^{(1)}),\cdots, \tilde{q}_n(\theta^{(M)})$ as in (\ref{eq_08})
and $\widehat{\phi}_n$ as in (\ref{eq_AbsErr}). If $\widehat{\phi}_{n=1}<\tau$, go to Step 4. Otherwise increase $n=n+1$
and update $\tilde{q}_n(\theta)$ and $\widehat{\phi}_n(\theta)$ until $\widehat{\phi}_n<\tau$.
\end{description}

\item[4] Calculate $\tilde{q}_n(\theta^{(M+1)}),\ldots,\tilde{q}_n(\theta^{(T)})$ and replace
$q(\theta^{(1)}),\ldots,q(\theta^{(T)})$ with $\tilde{q}(\theta^{(1)}),\ldots,\tilde{q}(\theta^{(T)})$ in
(\ref{ImpSamp}) to estimate $\widehat{\mathfrak{E}}$.

\end{description}
}}

\noindent In Step 1., we used the method by \citet{jasra:holmes:stephens:2005}, even though alternatives implemented in the {\sf label.switching} package of \cite{papastamoulis:illiopoulos:2010} or in \cite{rodriguez:walker:2014} could be implemented as well. The total number of $h$ values that are computed is $Tk!$ in the standard dual importance sampling scheme but decreases to $(Mk!) +|\mathfrak{A}(k)|(T-M)$
when using $\tilde{q}(\theta)$. The relative gain in the total number of terms is thus
\begin{equation}
\Delta(\mathfrak{A}(k)) = \dfrac{(Mk! )+  |\mathfrak{A}(k)|(T-M)}{T
k!}=\dfrac{M}{T}\left(1-\dfrac{|\mathfrak{A}(k)|}{k!}\right)+\dfrac{|\mathfrak{A}(k)|}{k!} ~. \label{eq_Delta}
\end{equation}
The gain will thus depend on how small $|\mathfrak{A}(k)|$ is, when compared with $k!$, hence ultimately on the acceptable mean
absolute difference $\tau$.

\section{Simulation study}

\begin{figure}[ht!]
\begin{center}
\setlength{\unitlength}{1cm}
\begin{picture}(14,3.5)
\put(0,0){\includegraphics[width=7cm, height=3.5cm]{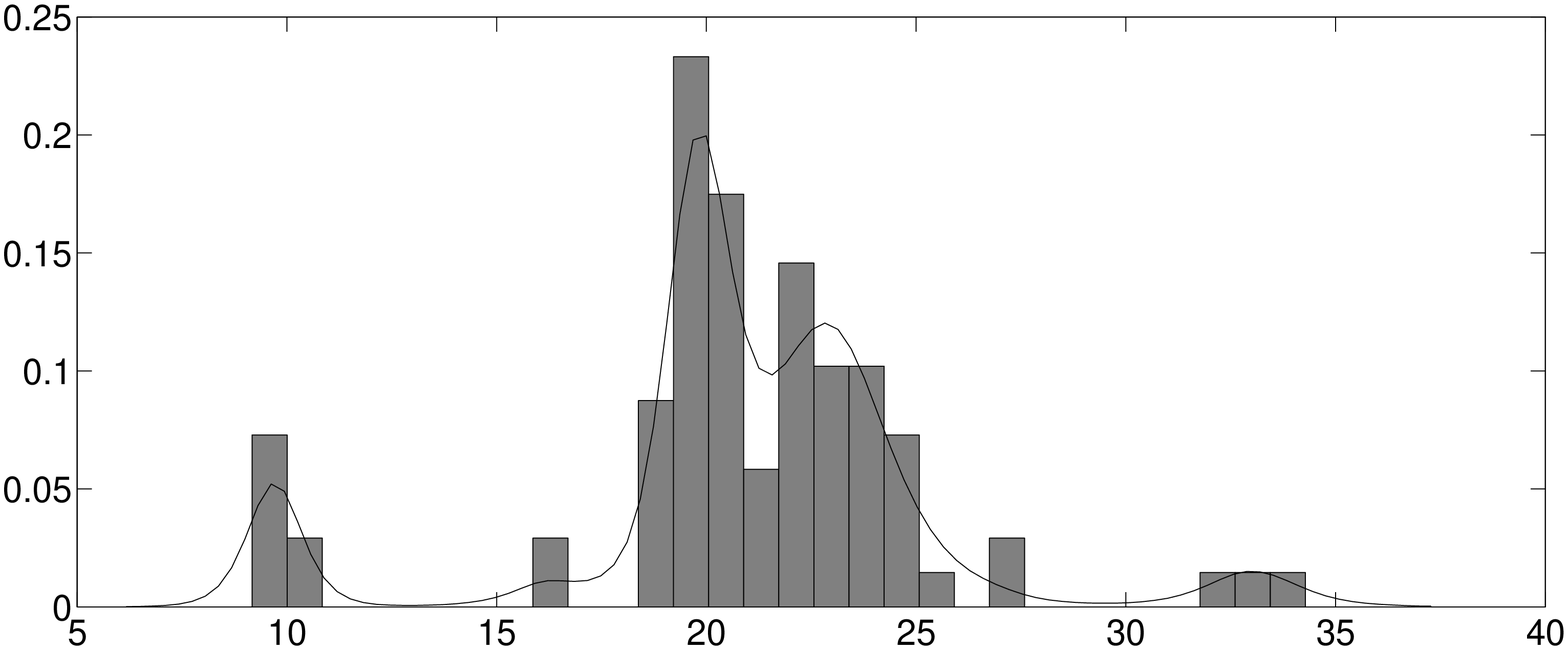}}
\put(7,0){\includegraphics[width=7cm, height=3.5cm]{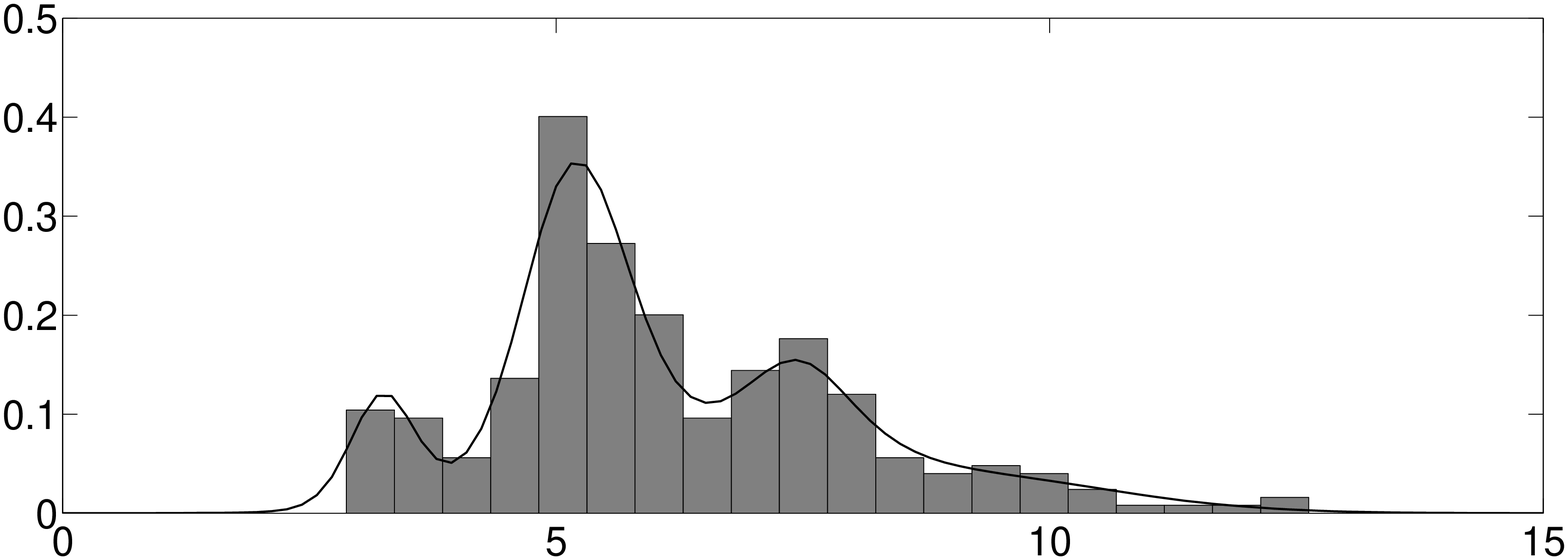}}
\rput(3.5,-0.1){$x$-values} \rput(10.5,-0.1){$x$-values}
\rput(5.9,3){(a} \rput(12.9,3){(b}
\end{picture} \caption{Histogram of the data against estimated six- and four- Gaussian mixture densities (solid line) 
for (a) the Galaxy dataset and (b) the fishery dataset, respectively.}\label{Hist_Data}
\end{center}\end{figure}

Two simulated mixture datasets and two real datasets are used to examine the
performance of seven marginal likelihood estimators. The simulated datasets, $D_1$ and $D_2$, are; 
\begin{itemize}
\item $D_1 : x_1,\dots,x_{60} \sim 0.3N(-1,1)+0.7N(5,2^2)$
\item $D_2 : x_1,\dots,x_{80} \sim 0.15N(-5,1) +0.65N(1,2^2) + 0.2N(6,1)$
\end{itemize}
where $N(5,2^2)$ denotes a normal distribution with a mean of 5 and a standard deviation of 2.  Two real datasets,
called galaxy and fishery datasets respectively, are shown in Figure \ref{Hist_Data}. They have been frequently used in
the literature as benchmarks \citep[see,
e.g.][]{chib:1995,fruhwirth:2006,jasra:holmes:stephens:2005,richardson:green:1997,stephens:2000}. 

\noindent Gaussian and Dirichlet priors are used for the means $\{\mu_i\}^k_{i=1}$ and proportions $\blambda$,
\[
\{\mu_i\}^k_{i=1} \sim N(0,10^2) \hspace{0.5cm} {\rm and} \hspace{0.5cm} (\lambda_1,\ldots,\lambda_k) \sim \text{Dir}(1,\ldots, 1) ~.
\]
For the variance parameters $\{\sigma_i^2\}^k_{i=1}$, inverse Gamma distributions with two sets of hyperparameters, $IG(2,3)$ and $IG(2,15)$, are considered. With the second calibration, label switching naturally occurred in Gibbs sequences in our simulation experiments. Removing the first 5000 Gibbs simulations as burn-ins, $10^4$ Gibbs simulations are used to approximate $\mathfrak{E}(k)$.

\noindent Firstly, a sensitivity analysis is conducted about the expected relative contribution of $h_{\sigma_i}$ to
$q(\theta)$ with respect to $M$. Then we set the values for both $M$ and $\tau$. In Section \ref{sub:simults}, the
performance of seven estimators for $\mathfrak{E}(k)$ are compared through a large simulation study, which confirms that
the asymptotic variance of $\widehat{\mathfrak{E}}(k)$ based on (\ref{eq_DS}) is smaller than when based on
(\ref{eq_perms_q}).
 
\subsection{Determining $M$ and $\tau$}

The approximation set is constructed in two steps. First, we compute
$\widehat{\mathbb{E}}_{h_{\sigma_c}}[\eta_{\sigma_1(\theta)}]$, $\ldots$,
$\widehat{\mathbb{E}}_{h_{\sigma_c}}[\eta_{\sigma_{k!}(\theta)}]$, based on reduced samples of size $M$ as in
(\ref{eq_eta}). Second, we derive which terms are negligible when compared with the threshold $\tau$. In our
experiments, we chose $\tau$ conservatively so that all zero terms are
identified. In MatLab, $10^{-324}$ is rounded down to $0$ thus $\tau=10^{-324}$ was chosen for the following simulation
studies.

\noindent The expected relative contribution measures for $D_1$ and $D_2$ are given in Tables \ref{Q_D1} and \ref{Q_D2}, respectively. For $J=10^2$ initial Gibbs simulations, significantly contributing clusters are easily identified by $\{\widehat{\mathbb{E}}_{h_{\sigma_1}}[\eta_{\sigma_i}(\theta)]\}_{i=1}^{k!}$ and both $|\mathfrak{A}(k)|$ and $\widehat{\phi}$ are relatively stable against $M$. Under a natural lack of label switching, $q(\theta)$ seems to be well approximated using only $h_{\sigma_1}(\theta)$, as seen in Table \ref{Q_D1}. Even when some label switching occurs in a Gibbs sequence corresponding to a Gaussian mixture model with three components, only two terms, $h_{\sigma_1}(\theta)$ and $h_{\sigma_2}(\theta)$, significantly contribute to $q(\theta)$, as seen in Table \ref{Q_D2}. For the subsequent analyses in this paper, we chose
$J=10^2$, $M=10^3$ and $\tau=10^{-324}$. \\

\begin{table} [ht!]\small
$$\begin{array}{c | c c c}
M &  \{\widehat{\mathbb{E}}_{h_{\sigma_1}}[\eta_{\sigma_i}(\theta)]\}_{i=1}^{k!} & |\mathfrak{A}(k)| & \widehat{\phi}  \\ \hline
10^2 & [1, 1.89\times 10^{-102}] & 1 & 0 \\
10^3 & [1, 5.25 \times 10^{-90}] & 1 & 0 \\
10^4 & [1, 4.62\times 10^{-91}] & 1 & 0 \\  
10^5 & [1, 3.56\times 10^{-80}] & 1 & 0 \\  \hline
\end{array} $$
\caption{ Estimates for $\{\widehat{\mathbb{E}}_{h_{\sigma_1}}[\eta_{\sigma_i}]\}_{i=1}^{k!}$, 
$|\mathfrak{A}(k)|$ and $\widehat{\phi}$ against $M$ for $D_1$ ($k=2$). The prior for a variance 
parameter is $IG(2,3)$. Note that due to rounding errors, the sum of the contribution ratios does not equal one.}
\label{Q_D1} \end{table}

\begin{table} [ht!]\small
$$ \begin{array}{c c | c c c}
 M &  \{\widehat{\mathbb{E}}_{h_{\sigma_1}}[\eta_{\sigma_i}]\}_{i=1}^{k!} & |\mathfrak{A}(k)| & \widehat{\phi} \\ \hline
 10^2 & [3.56\times 10^{-16},9.53\times 10^{-160}, 5.05\times 10^{-55}, 8.27\times 10^{-144},1.0, 4.64\times 10^{-65}] & 2 & 0 \\
 10^3 & [1.22\times 10^{-8}, 1.11\times 10^{-144}, 3.01\times 10^{-49}, 3.08\times 10^{-125}, 1.0, 2.27\times 10^{-53}] & 2 & 0 \\
 10^4 & [2.03\times 10^{-8},8.31\times 10^{-136}, 1.76\times 10^{-43}, 2.61\times 10^{-95}, 1.0, 4.87\times 10^{-49}] & 2 & 0 \\ 
 10^5 & [1.04\times 10^{-5}, 1.56\times 10^{-122}, 1.51 \times 10^{-44}, 4.30\times 10^{-87}, 1.0, 2.27\times 10^{-39}] & 2 & 0 \\ \hline
\end{array} 
$$ 
\caption{ Estimates for $\{\widehat{\mathbb{E}}_{h_{\sigma_1}}[\eta_i]\}_{i=1}^{k!}$, $|\mathfrak{A}(k)|$ and $\widehat{\phi}$ with respect to $M$ for $D_2$ ($k=3$). The prior for a variance parameter is $IG(2,15)$. Note that due to rounding errors, the sum of the contribution ratios does not equal one. } \label{Q_D2} 
\end{table}

\newpage

\subsection{Simulation results}\label{sub:simults}

The following seven marginal likelihood estimators using an equal number of proposals are compared;

\begin{description}
\item [$\widehat{\mathfrak{E}}^*_{Ch}$ :] Chib's method (\ref{eq_01}) using $T=10^4$ samples and multiplying by $k!$
to compensate for a lack of label switching; 
\item [$\widehat{\mathfrak{E}}_{Ch}$ :] Chib's method with density estimate (\ref{eq_01}), using  $T=10^4$ randomly permuted Gibbs samples;
\item [$\widehat{\mathfrak{E}}_{IS}$ :] Importance sampling using $q$ as in (\ref{eq_q_mle}), with a maximum likelihood
estimate for $z_1^o,\dots, z_n^o$ and $T=10^4$ particles;
\item [$\widehat{\mathfrak{E}}_{DS}$ :] Dual importance sampling using $q$ as in (\ref{eq_DS}), with $T=10^4$ particles
and $J=100$ Gibbs samples in $q(\theta)$; 
\item [$\widehat{\mathfrak{E}}^{A}_{DS}$ :] Dual importance sampling using an approximation as in (\ref{eq_08}), with $T=10^4$ particles, $J=100$ and $M=10^3$;
\item [$\widehat{\mathfrak{E}}_{J_1}$ :] Importance sampling using $q$ as in (\ref{eq_perms_q}), with $T=10^4$
particles. When $k<6$, $J_1=100k!$ and otherwise $J_1=5000$;  
\item [$\widehat{\mathfrak{E}}_{\small BS}$ :] Bridge sampling (\ref{eq_BS}), using $M_1=M_2=6\times 10^3$ samples and $q(\theta)$ as in (\ref{eq_perms_q}) via 10 iterations. For $q$, it is set as  $J_1=4000$ and label switching is imposed in hyperparameters $\{\theta^{(j)},z^{(j)}\}_{j=1}^{J_1}$.
\end{description}

\noindent The marginal likelihood estimates (in log scales) and the effective sample size (ESS) ratios, $R={\rm ESS}/T$,
are summarized in Figures \ref{Sim_1} and \ref{Sim_2} by boxplots, based on 50 replicates. Subscripts of
$\widehat{\mathfrak{E}}$ and $R$ denote the estimating technique. Note that a
modified ESS, provided by equation (35) in \citet{doucet:godwill:andrieu:2000}, is used here for numerical stability.
All estimators are based on $10^4$ proposals, as in Table \ref{table_computation}, where summing up
the second and third columns leads to a fixed total number of function evaluations. Within our setup,
$\widehat{\mathfrak{E}}_{\small IS}$ is the least demanding in terms of computational workload while the remaining importance estimators require the same computing time, except for $\widehat{\mathfrak{E}}^{A}_{DS}$.

\begin{table} \[\begin{array}{c |c c | c}
{\rm Estimate} & \text{Number of posterior} & \text{Number of marginal posterior}  & \text{Number of proposals} \\
& \text{evaluations} & \text{density evaluations in $q$} &  \\ \hline
\widehat{\mathfrak{E}}_{IS} & T & Tk! & T \\ 
\widehat{\mathfrak{E}}_{DS} & T & TJk! & T  \\
\widehat{\mathfrak{E}}^{A}_{DS} & T & (M + (T-M)|\mathfrak{A}(k)|/k!)Jk! & T \\
\widehat{\mathfrak{E}}_{J_1} & T & TJ_1 & T \\
\widehat{\mathfrak{E}}_{BS} & M_1 & (M_1+M_2)J_1 & M_1+M_2 \\ \hline
\end{array} \]
\caption{Computation steps required by different evidence estimation approaches. Note that the required computation for
$\widehat{\mathfrak{E}}_{\small BS}$ is given per iteration.}  \label{table_computation}
\end{table}

\subsubsection{Simulated mixture dataset}

\begin{figure}[ht!]
\begin{center}
\setlength{\unitlength}{1cm}
\begin{picture}(14.5,6)
\put(0,3){\includegraphics[width=9cm, height=3cm]{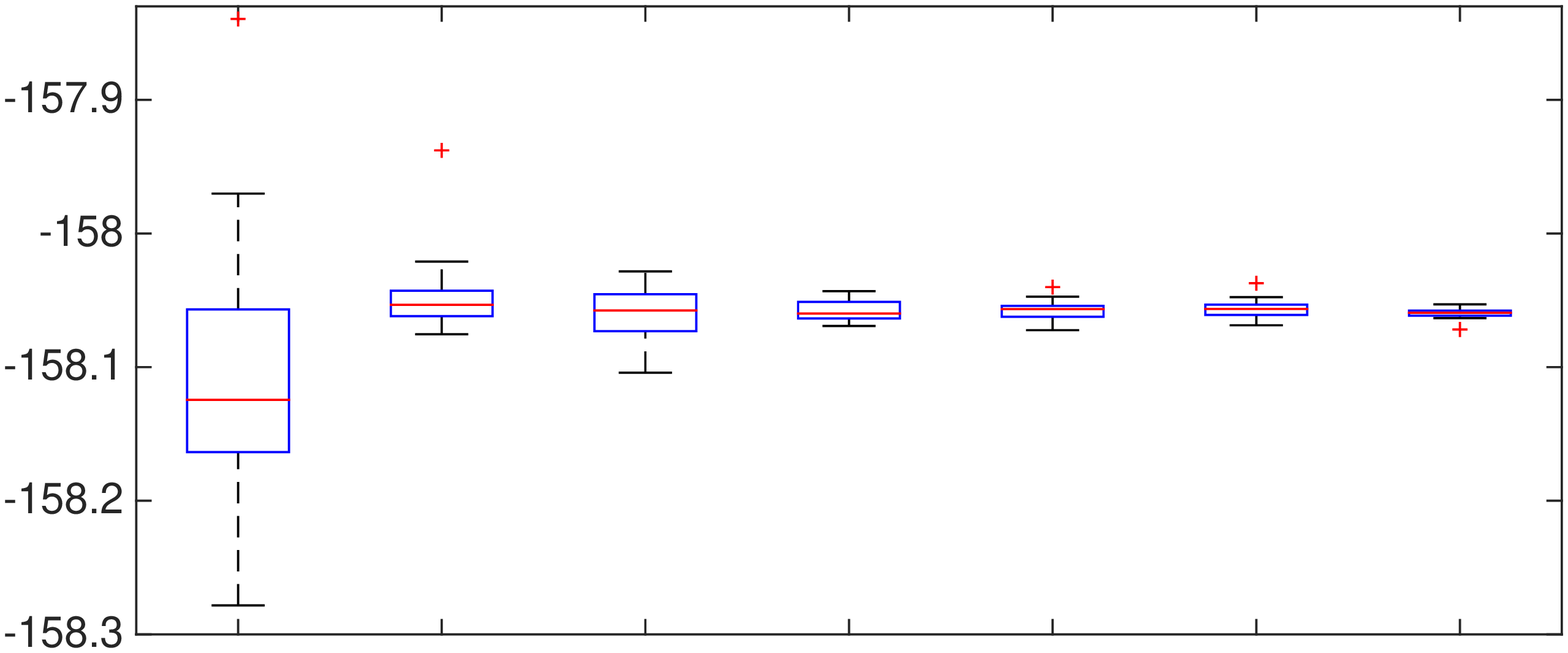}}
\put(0,0){\includegraphics[width=9cm, height=3cm]{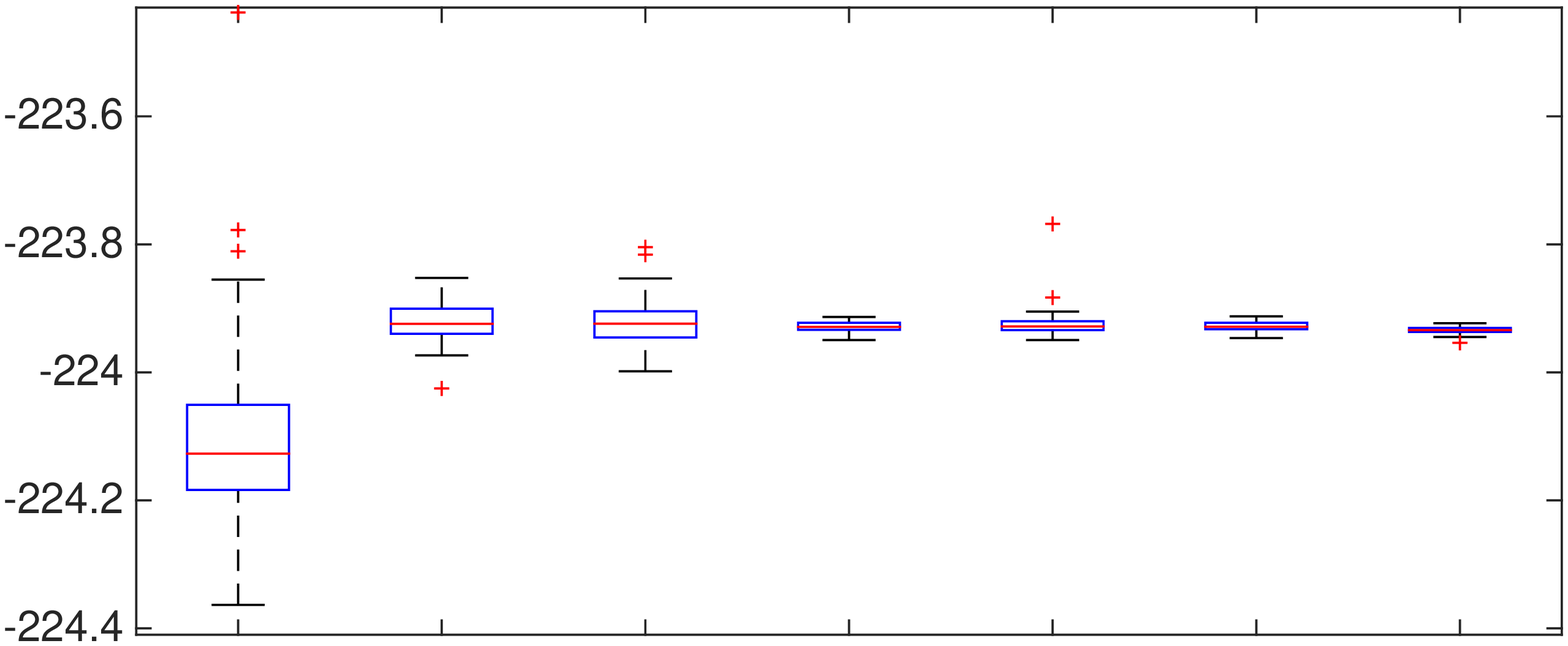}}

\put(8.5,3){\includegraphics[width=4.5cm, height=3cm]{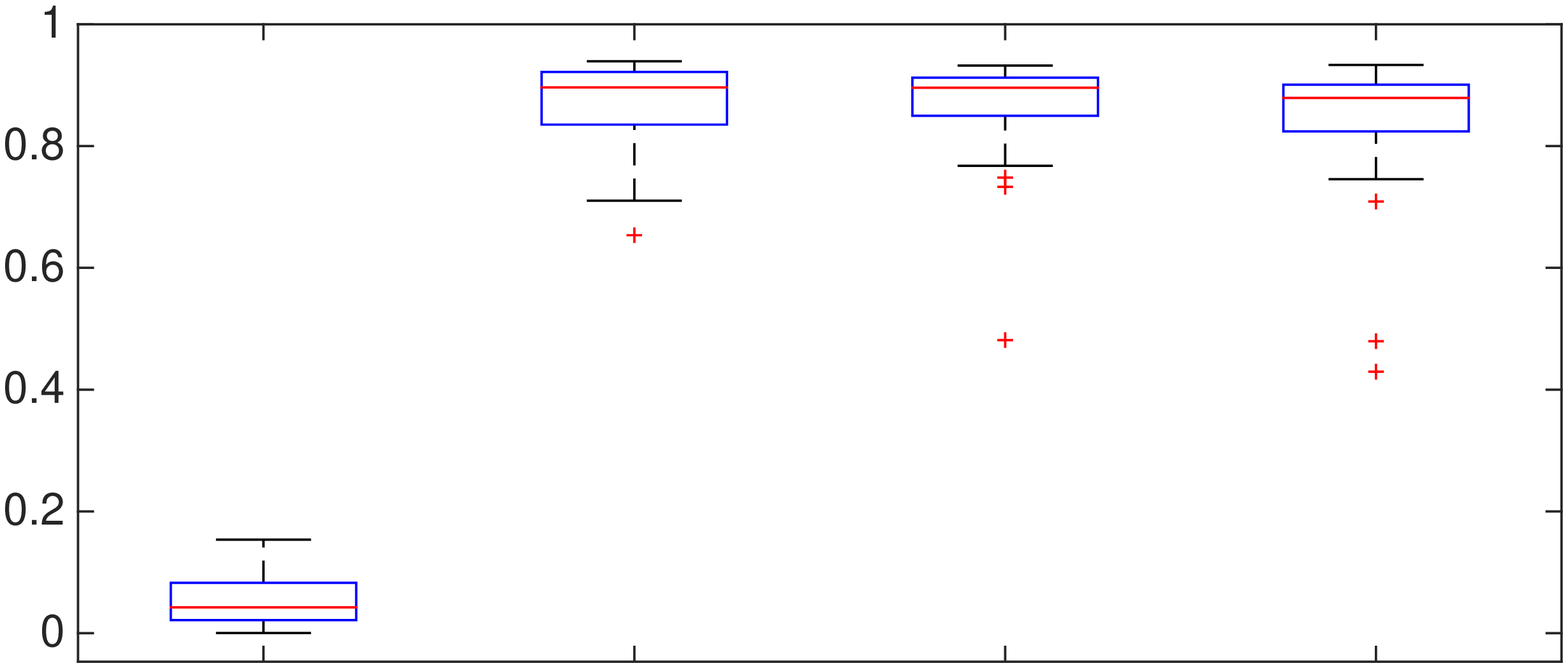}}
\put(8.5,0){\includegraphics[width=4.5cm, height=3cm]{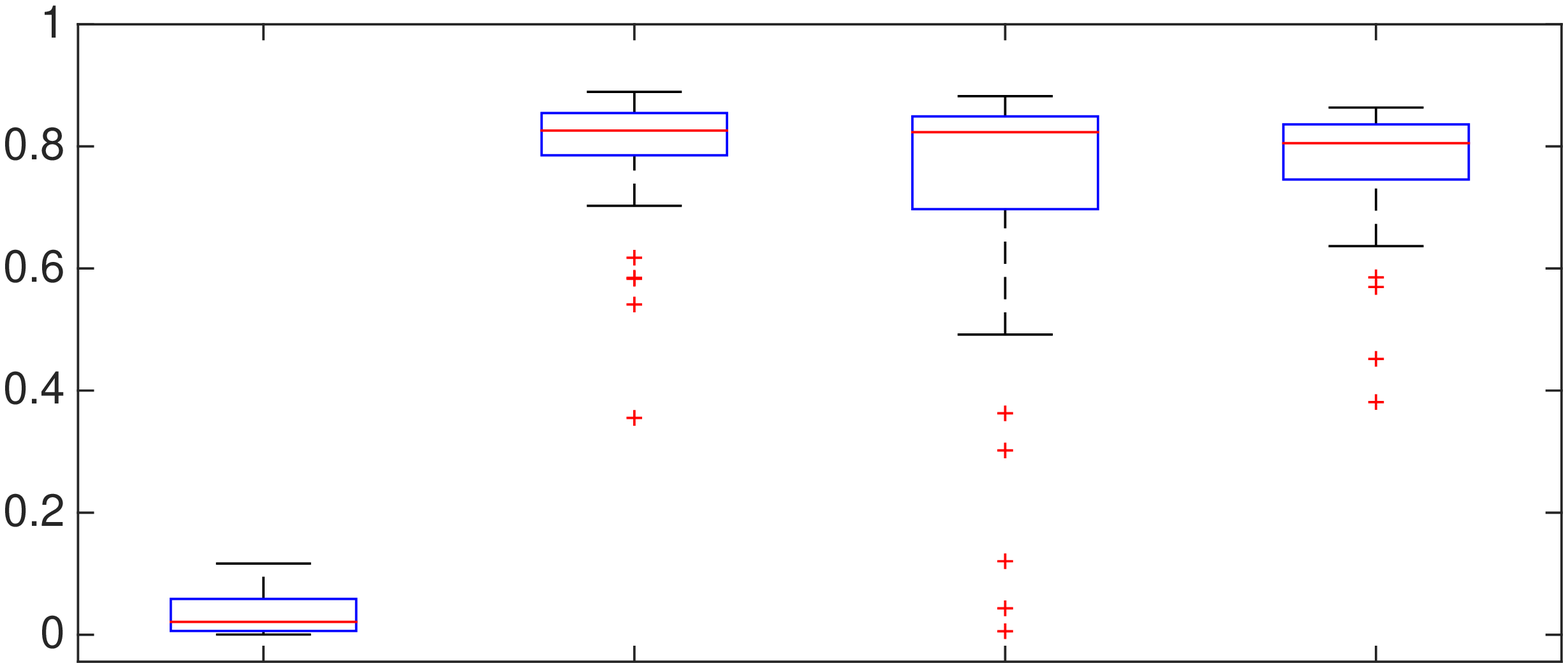}}

\rput(1.7,0.1){\footnotesize$\widehat{\mathfrak{E}}_{IS}$}  \rput(2.7,0.1){\footnotesize$\widehat{\mathfrak{E}}^*_{Ch}$} 
\rput(3.8,0.1){\footnotesize$\widehat{\mathfrak{E}}_{Ch}$} \rput(4.8,0.1){\footnotesize$\widehat{\mathfrak{E}}_{DS}$} 
\rput(5.8,0.1){\footnotesize$\widehat{\mathfrak{E}}^{A}_{DS}$} 
\rput(6.8,0.1){\footnotesize$\widehat{\mathfrak{E}}_{J_1}$}  \rput(7.8,0.1){\footnotesize$\widehat{\mathfrak{E}}_{BS}$}

\rput(1.7,3.1){\footnotesize$\widehat{\mathfrak{E}}_{IS}$}  \rput(2.7,3.1){\footnotesize$\widehat{\mathfrak{E}}^*_{Ch}$} 
\rput(3.8,3.1){\footnotesize$\widehat{\mathfrak{E}}_{Ch}$} \rput(4.8,3.1){\footnotesize$\widehat{\mathfrak{E}}_{DS}$} 
\rput(5.8,3.1){\footnotesize$\widehat{\mathfrak{E}}^{A}_{DS}$} 
\rput(6.8,3.1){\footnotesize$\widehat{\mathfrak{E}}_{J_1}$}  \rput(7.8,3.1){\footnotesize$\widehat{\mathfrak{E}}_{BS}$}

\rput(9.6,0.1){\footnotesize$R_{IS}$} \rput(10.4,0.1){\footnotesize$R_{DS}$} 
\rput(11.4,0.1){\footnotesize$R^{A}_{DS}$} \rput(12.3,0.1){\footnotesize$R_{J_1}$}

\rput(9.6,3.1){\footnotesize$R_{IS}$} \rput(10.4,3.1){\footnotesize$R_{DS}$} 
\rput(11.4,3.1){\footnotesize$R^{A}_{DS}$} \rput(12.3,3.1){\footnotesize$R_{J_1}$}

\end{picture} \end{center}
\caption{
Boxplots of evidence estimates in log scale {\em (left, middle)} and effective sample sizes ratios {\em
(right)}. Mixture models with two and three Gaussian components are fitted to {\em (top)}  $D_1$ and {\em (bottom)}
$D_2$, respectively. The prior for $\{\sigma_i^2\}^k_{i=1}$ is $IG(2,3)$ and label switching did not occur in Gibbs
samples. One outlier of $\widehat{\mathfrak{E}}_{IS}$ in the top-left panel is discarded. }  \label{Sim_1}
\end{figure}

\begin{figure}[ht!]
\begin{center}
\setlength{\unitlength}{1cm}
\begin{picture}(14.5,6)
\put(0,3){\includegraphics[width=9cm, height=3cm]{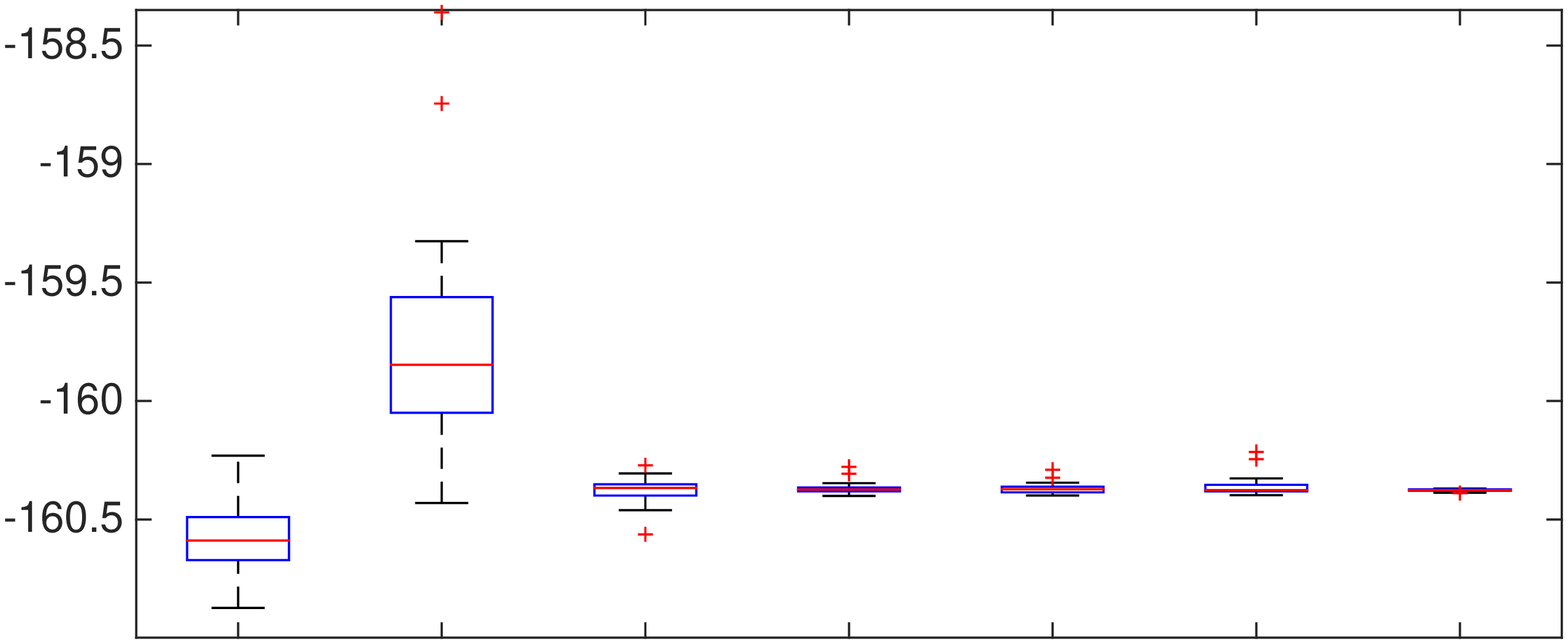}}
\put(0,0){\includegraphics[width=9cm, height=3cm]{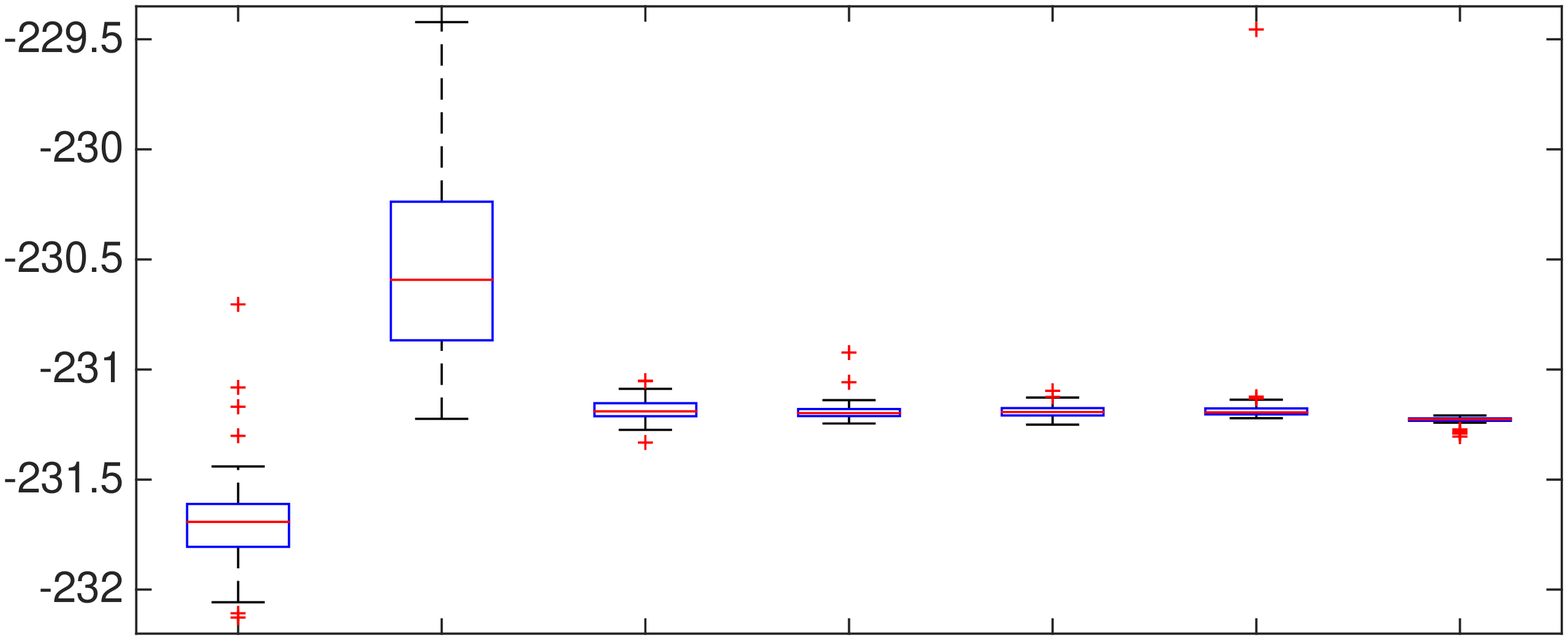}}

\put(8.5,3){\includegraphics[width=4.5cm, height=3cm]{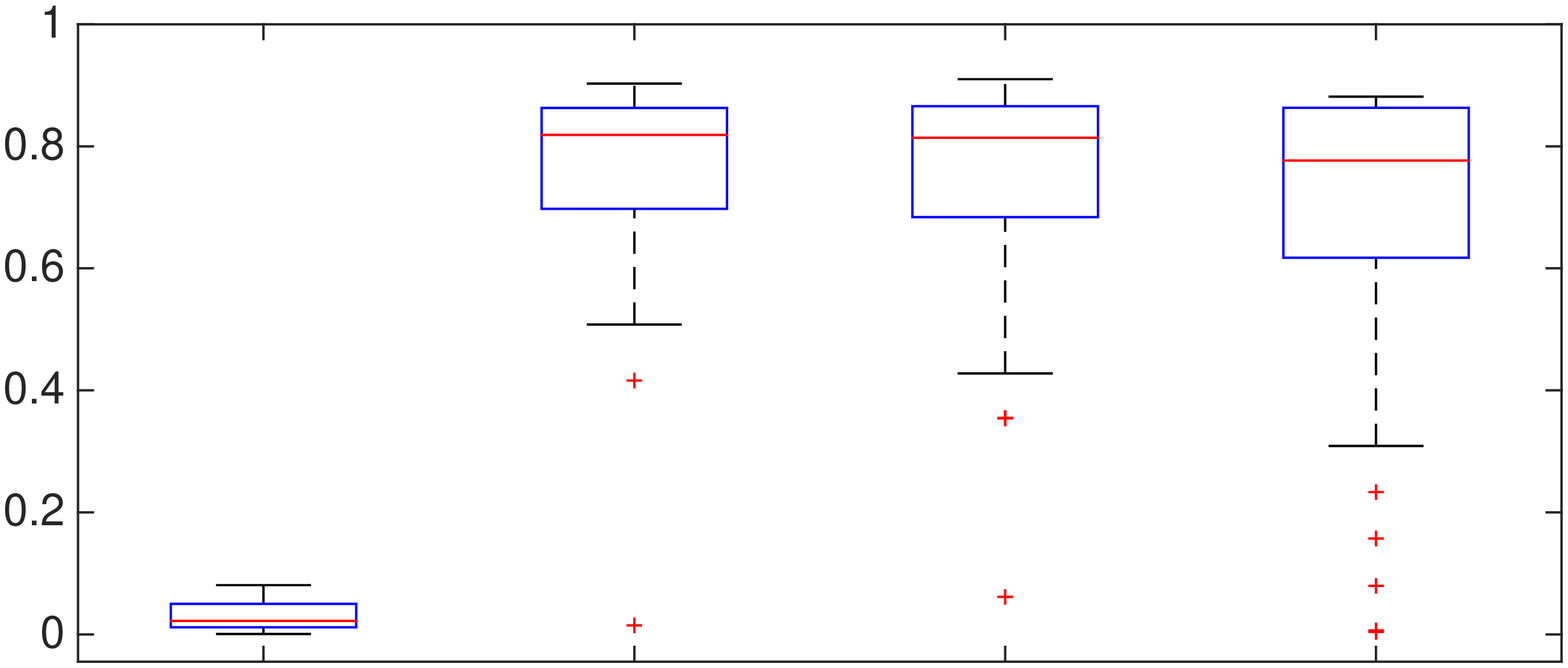}}
\put(8.5,0){\includegraphics[width=4.5cm, height=3cm]{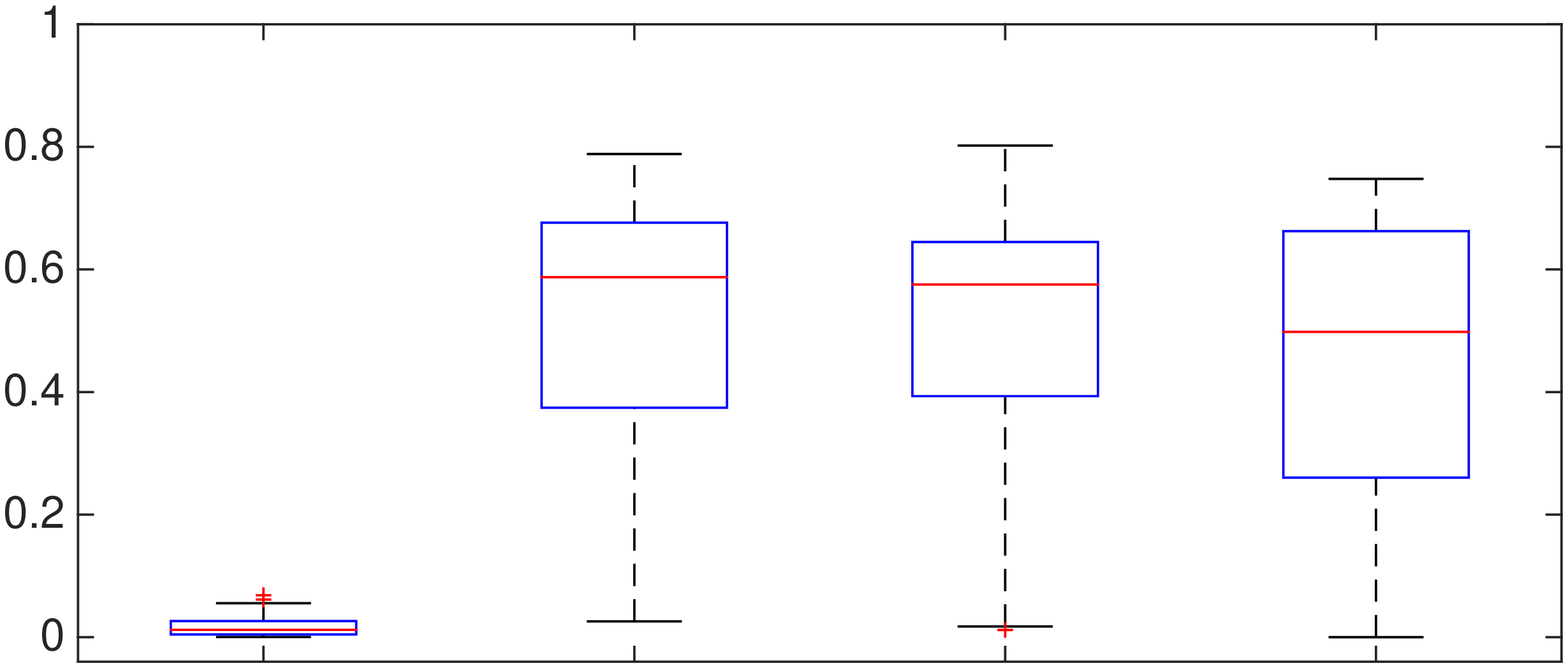}}

\rput(1.7,0.1){\footnotesize$\widehat{\mathfrak{E}}_{IS}$}  \rput(2.7,0.1){\footnotesize$\widehat{\mathfrak{E}}^*_{Ch}$} 
\rput(3.8,0.1){\footnotesize$\widehat{\mathfrak{E}}_{Ch}$} \rput(4.8,0.1){\footnotesize$\widehat{\mathfrak{E}}_{DS}$} 
\rput(5.8,0.1){\footnotesize$\widehat{\mathfrak{E}}^{A}_{DS}$} \rput(6.8,0.1){\footnotesize$\widehat{\mathfrak{E}}_{J_1}$} 
\rput(7.7,0.1){\footnotesize$\widehat{\mathfrak{E}}_{BS}$}

\rput(1.7,3.1){\footnotesize$\widehat{\mathfrak{E}}_{IS}$}  \rput(2.7,3.1){\footnotesize$\widehat{\mathfrak{E}}^*_{Ch}$} 
\rput(3.8,3.1){\footnotesize$\widehat{\mathfrak{E}}_{Ch}$} \rput(4.8,3.1){\footnotesize$\widehat{\mathfrak{E}}_{DS}$} 
\rput(5.8,3.1){\footnotesize$\widehat{\mathfrak{E}}^{A}_{DS}$} \rput(6.8,3.1){\footnotesize$\widehat{\mathfrak{E}}_{J_1}$} 
\rput(7.7,3.1){\footnotesize$\widehat{\mathfrak{E}}_{BS}$}

\rput(9.5,0.1){\footnotesize$R_{IS}$} \rput(10.4,0.1){\footnotesize$R_{DS}$} 
\rput(11.3,0.1){\footnotesize$R^{A}_{DS}$} \rput(12.1,0.1){\footnotesize$R_{J_1}$} 

\rput(9.5,3.1){\footnotesize$R_{IS}$} \rput(10.4,3.1){\footnotesize$R_{DS}$} 
\rput(11.3,3.1){\footnotesize$R^{A}_{DS}$} \rput(12.1,3.1){\footnotesize$R_{J_1}$} 

\end{picture} \end{center}
\caption{
Boxplots of evidence estimates in log scale {\em (left, middle)} and effective sample sizes ratios {\em
(right)}. Mixture models with two and three Gaussian components are fitted to {\em (top)} $D_1$ and {\em (bottom)}
$D_2$, respectively. The prior for $\{\sigma_i^2\}^k_{i=1}$ is $IG(2,15)$ and label switching naturally occurred in
Gibbs samples. Two outliers for $\widehat{\mathfrak{E}}^*_{Ch}$ in the top-left panel are discarded.} \label{Sim_2}
\end{figure}

Mixture models of two and three components are fitted to $D_1$ and $D_2$ respectively. Regardless of the presence or not
of label switching in the resulting Gibbs sequences, all estimates based on importance sampling except
$\widehat{\mathfrak{E}}_{IS}$ coincide with $\widehat{\mathfrak{E}}_{Ch}$, albeit with smaller Monte Carlo variations as
seen in Figures \ref{Sim_1} and \ref{Sim_2}. When a suitable approximate for $q(\theta)$ is used for the dual importance
sampling, no significant difference in the estimates $\log(\widehat{\mathfrak{E}}(k))$ and in the effective sample sizes
are observed. The mean sizes of $\mathfrak{A}(k)$ in Table \ref{table_data} are always smaller than $k!$ and it shows
that $\mathfrak{E}(k)$ can be estimated with a lesser computational workload. When posterior modes are very well
separated (no natural label switching ever present in Gibb sequences), the number of evaluations in $q$ is reduced
almost by the maximal factor of $1/k!$. In Table \ref{table_time_data}, the least computational demand is observed for
the chib's methods while the bridge sampling costs more than 100 times. When $\mathfrak{A}(k) < k!$, some reduction in CPU time for $\widehat{\mathfrak{E}}(k)_{DS}^A$ is observed due to ignoring zero function evaluation. 

\noindent Disagreement in the values of $\widehat{\mathfrak{E}}_{IS}$ versus $\widehat{\mathfrak{E}}_{Ch}$ shows that an
importance function may fail to properly approximate $p_k(\bx|\theta)\pi(\theta)$, resulting in an unreliable estimate
with large variation. Significantly small effective sample sizes (i.e., very small values for $R_{IS}$) back this
observation. In our simulation experiments, we observed that $\widehat{\mathfrak{E}}_{BS}$ is correctly calibrated for a
large  value of $J_1$ (i.e., a large number of conditional densities in $q$). When label switching naturally occurs, as
in the Gibbs sequence under the variance prior $IG(2,15)$, $\widehat{\mathfrak{E}}^*_{Ch}$ disagrees with the other
estimates, see Figure \ref{Sim_2}. Unsurprisingly, this indicates that the simplistic correction through a
multiplication by $k!$ is of no use, as reported in \citet{neal:1999}, \cite{fruhwirth:2006} and \citet{marin:robert:2008}.

\begin{table}[ht!]
\[ \begin{array} {c c c | c c | c c }
 D & k & k! & |\mathfrak{A}_1(k)| & \Delta(\mathfrak{A}_1) & |\mathfrak{A}_2(k)| & \Delta(\mathfrak{A}_2)  \\  \hline
D_1 & 2 & 2 & 1.00 ~(0.00) & 0.55 ~(2.26\times 10^{-16})& 1.73 ~(0.45) & 0.88 ~(0.20) \\ 
D_2 & 3 & 6 & 1.02 ~(0.14) & 0.25 ~(0.02) & 2.18 ~(0.60) & 0.43 ~(0.09) \\
\hline
\end{array} \]
\caption{
Mean and standard deviation \emph{(values in brackets)} estimates for the approximation set size, $|\mathfrak{A}(k)|$,
and the reduction rate of a number of evaluated $h$-terms, $\Delta(\mathfrak{A})$, as in (15) for $D_1$ and $D_2$.
Subscripts 1 and 2 indicate the results using the priors $\sigma^2\sim IG(2,3)$ and $\sigma^2\sim IG(2,15)$,
respectively. } \label{table_data}
\end{table}

\begin{table}[ht!]
\[ \begin{array} {| c | c c | c c |}
\hline 
\text{Estimator} & \multicolumn{2}{|c|}{D_1} & \multicolumn{2}{|c|}{D_2} \\  
 & \text{CPU}_1 & \text{CPU}_2 & \text{CPU}_1 & \text{CPU}_2 \\ \hline
\widehat{\mathfrak{E}}^*_{Ch} & 0.80 & 0.76 & 1.17 & 1.39 \\
\widehat{\mathfrak{E}}_{Ch} & 0.79 & 0.81 & 1.32 & 1.36 \\ 
\widehat{\mathfrak{E}}_{IS} & 2.54 & 2.65 & 3.35 & 3.35 \\
\widehat{\mathfrak{E}}_{DS} & 3.07 & 2.96 & 6.12 & 6.02 \\
\widehat{\mathfrak{E}}^{A}_{DS} & 2.87 & 3.07 & 3.77 & 5.47 \\
\widehat{\mathfrak{E}}_{J_1} & 2.42 & 3.34 & 6.02 & 6.88 \\
\widehat{\mathfrak{E}}_{BS} & 1.18\times 10^3 & 1.19\times 10^3 & 2.78\times 10^3 & 3.37\times 10^3 \\ \hline
\end{array} \]
\caption{
Elapsed CPU time in seconds for evidences approximation of mixture models for $D_1$ and $D_2$. Subscripts 1 and 2 of CPU
indicate the results using the priors $\sigma^2\sim IG(2,3)$ and $\sigma^2\sim IG(2,15)$, respectively. }
\label{table_time_data}
\end{table}

\subsubsection{Galaxy and fishery dataset}

\begin{figure}[ht!]
\begin{center}
\setlength{\unitlength}{1cm}
\begin{picture}(14,6)
\put(0,3){\includegraphics[width=9cm, height=3cm]{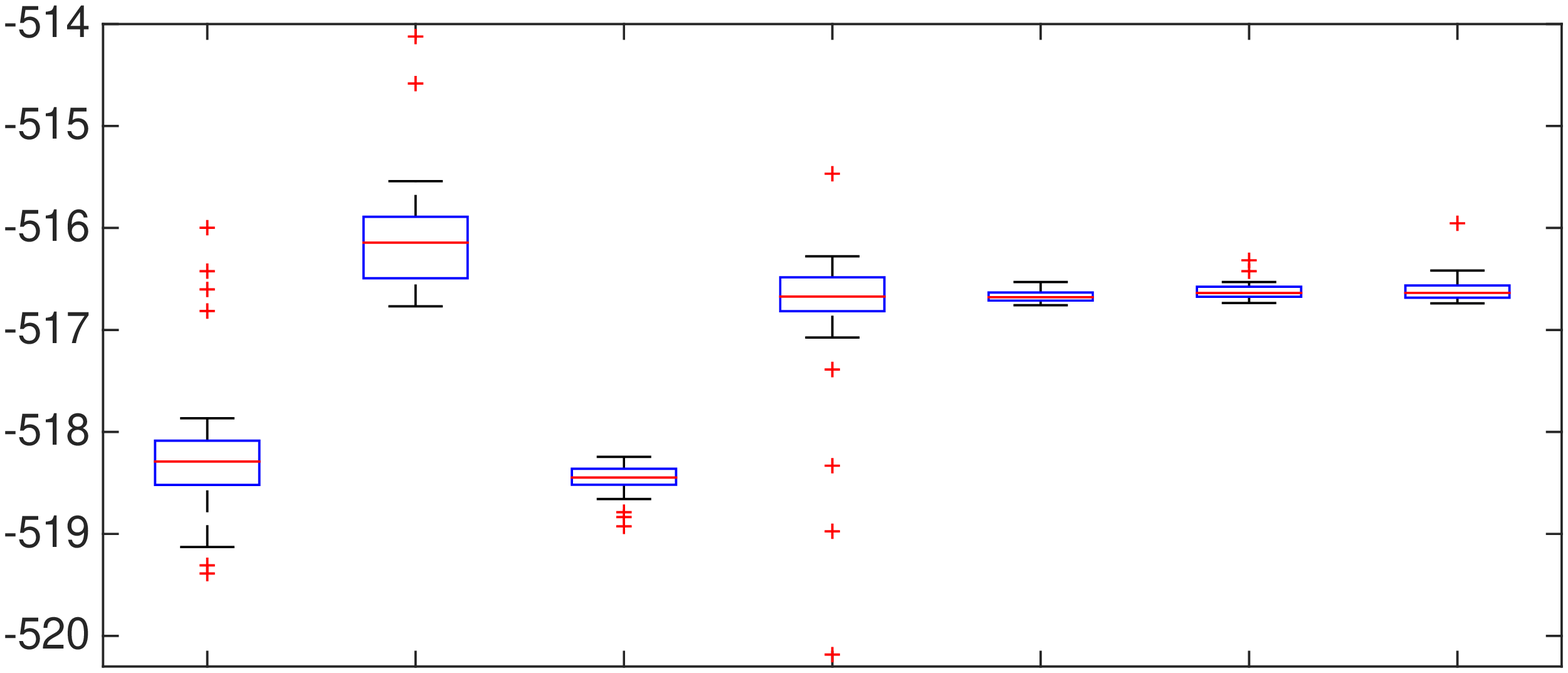}}
\put(0,0){\includegraphics[width=9cm, height=3cm]{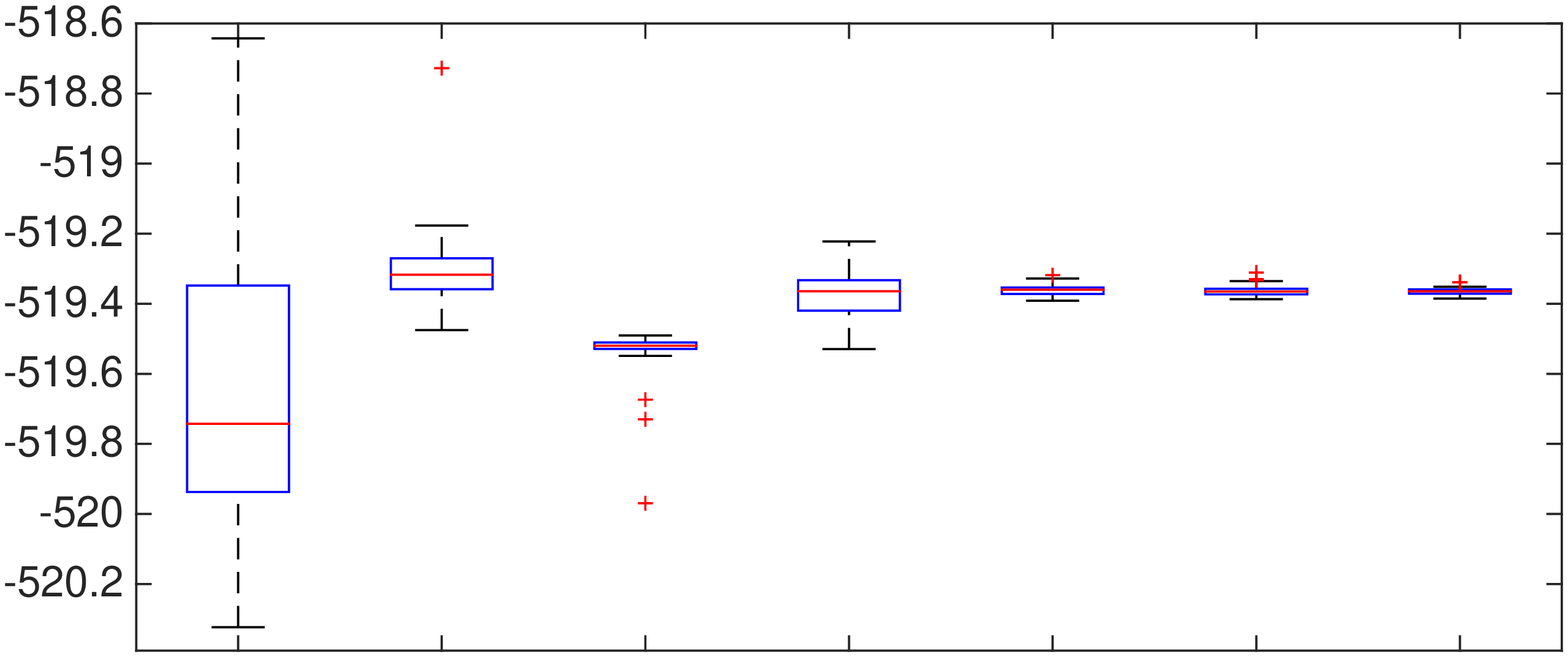}}
\put(8.5,3){\includegraphics[width=5cm, height=3cm]{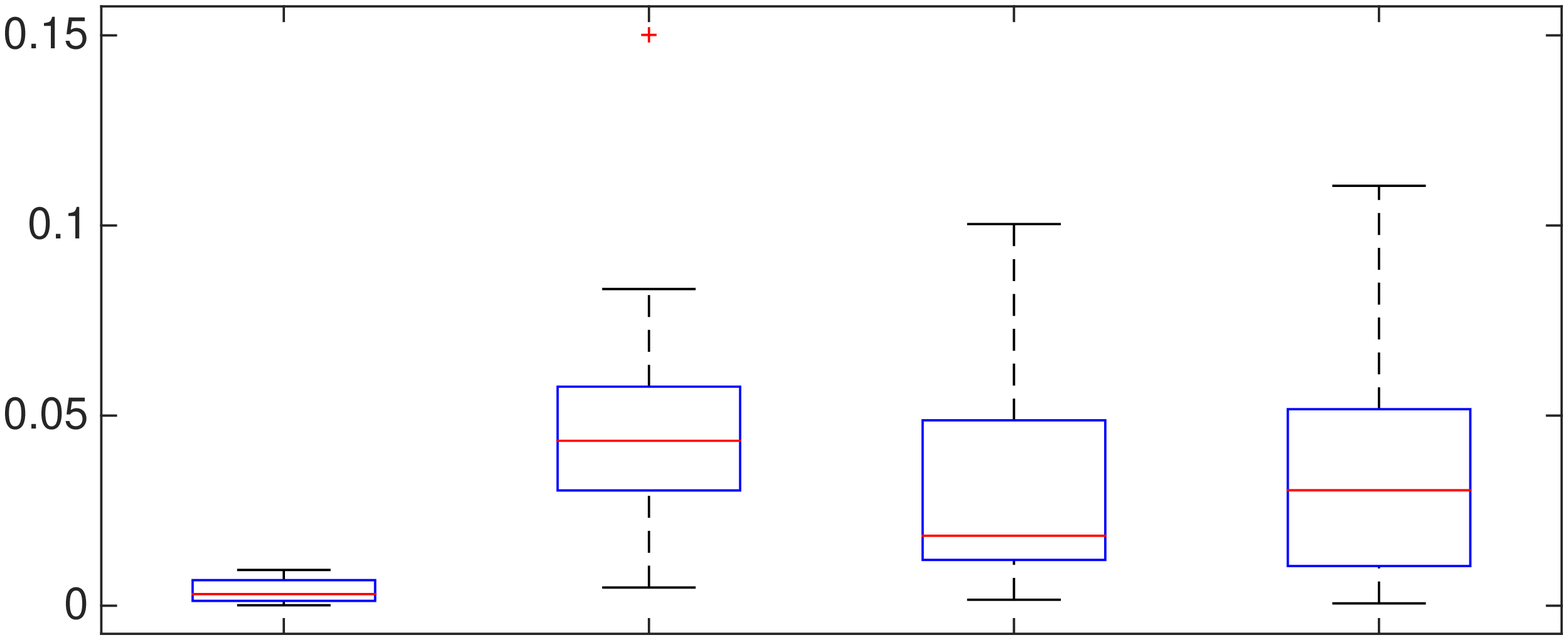}}
\put(8.5,0){\includegraphics[width=5cm, height=3cm]{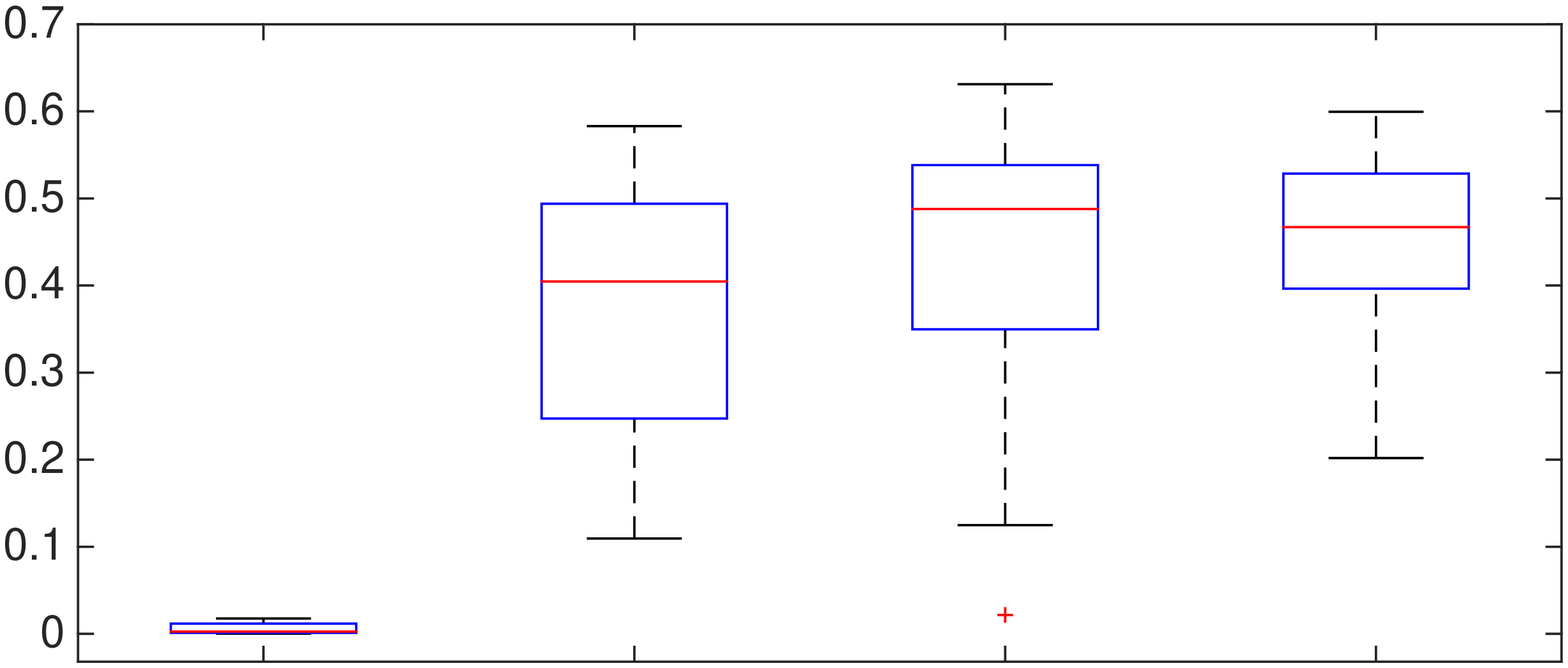}}

\rput(1.7,0.1){\footnotesize$\widehat{\mathfrak{E}}_{IS}$}  \rput(2.7,0.1){\footnotesize$\widehat{\mathfrak{E}}_{Ch}^*$} 
\rput(3.7,0.1){\footnotesize$\widehat{\mathfrak{E}}_{BS}$}  \rput(4.7,0.1){\footnotesize$\widehat{\mathfrak{E}}_{Ch}$}  
\rput(5.8,0.1){\footnotesize$\widehat{\mathfrak{E}}_{DS}$} \rput(6.7,0.1){\footnotesize$\widehat{\mathfrak{E}}^A_{DS}$} 
\rput(7.8,0.1){\footnotesize$\widehat{\mathfrak{E}}_{J_1}$}

\rput(1.7,3.1){\footnotesize$\widehat{\mathfrak{E}}_{IS}$}  \rput(2.7,3.1){\footnotesize$\widehat{\mathfrak{E}}_{Ch}^*$} 
\rput(3.7,3.1){\footnotesize$\widehat{\mathfrak{E}}_{BS}$}  \rput(4.7,3.1){\footnotesize$\widehat{\mathfrak{E}}_{Ch}$}  
\rput(5.8,3.1){\footnotesize$\widehat{\mathfrak{E}}_{DS}$} \rput(6.7,3.1){\footnotesize$\widehat{\mathfrak{E}}^A_{DS}$} 
\rput(7.8,3.1){\footnotesize$\widehat{\mathfrak{E}}_{J_1}$}

\rput(9.7,0.1){\footnotesize $R_{IS}$} \rput(10.7,0.1){\footnotesize$R_{DS}$} 
\rput(11.7,0.1){\footnotesize$R^{A}_{DS}$} \rput(12.6,0.1){\footnotesize$R_{J_!}$}

\rput(9.7,3.1){\footnotesize $R_{IS}$} \rput(10.7,3.1){\footnotesize$R_{DS}$} 
\rput(11.7,3.1){\footnotesize$R^{A}_{DS}$} \rput(12.6,3.1){\footnotesize$R_{J_1}$}

\end{picture} \end{center}
\caption{
Boxplots of evidence estimates in log scale {\em (left, middle)} and effective sample sizes ratios {\em (right)}. Mixture models with {\em (top)} three and {\em (bottom)} four Gaussian components are fitted to the fishery dataset. Two outliers of $\widehat{\mathfrak{E}}_{Ch}$ in the top-left panel are discarded.}  \label{Sim_fishery} \end{figure}

\begin{figure}[ht!]
\begin{center}
\setlength{\unitlength}{1cm}
\begin{picture}(14,9)
\put(0,6){\includegraphics[width=9cm, height=3cm]{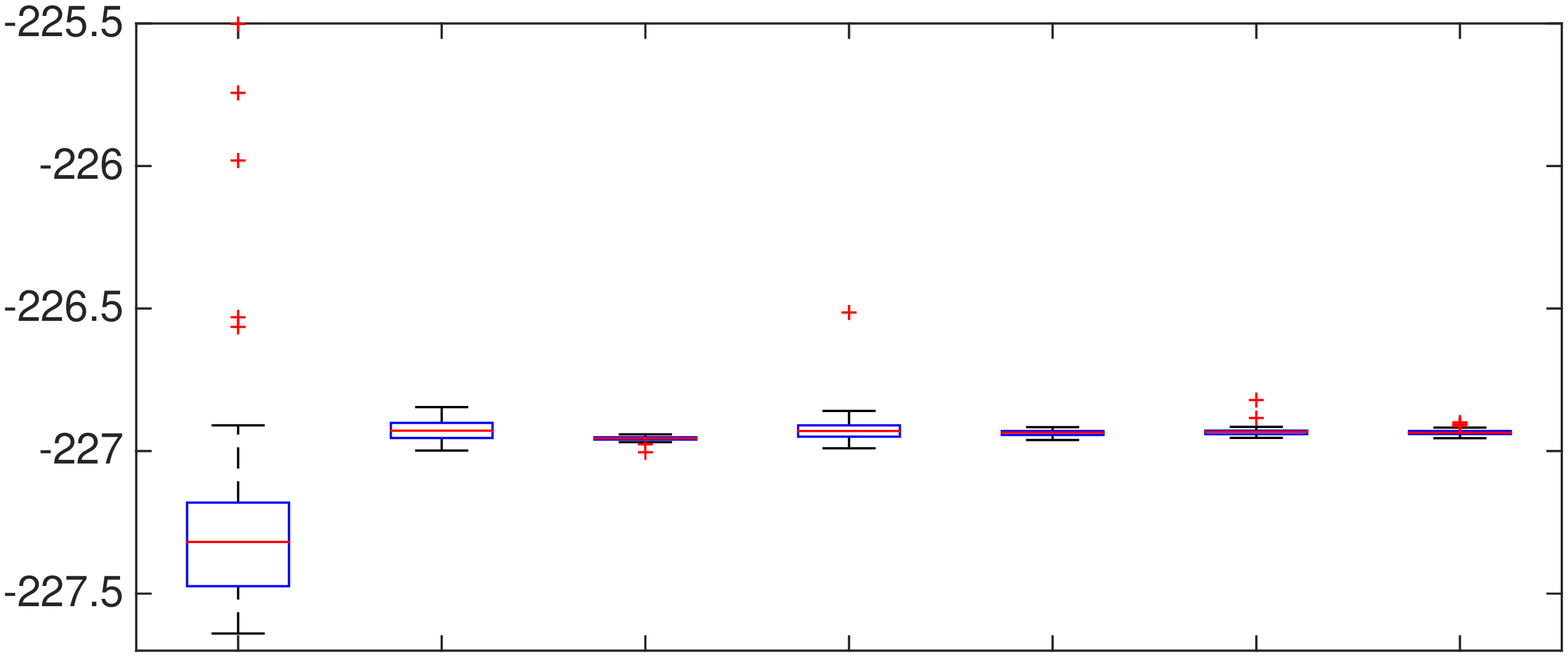}}
\put(0,3){\includegraphics[width=9cm, height=3cm]{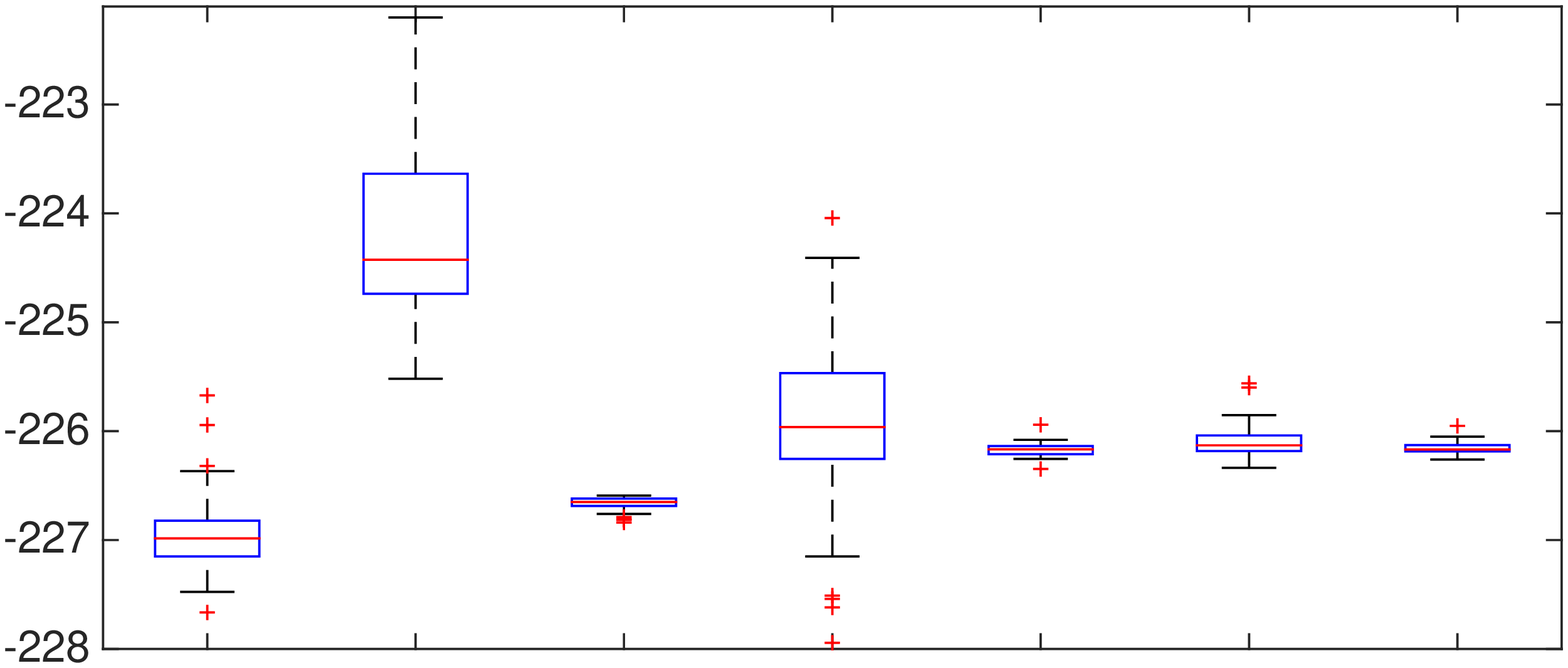}}
\put(0,0){\includegraphics[width=9cm, height=3cm]{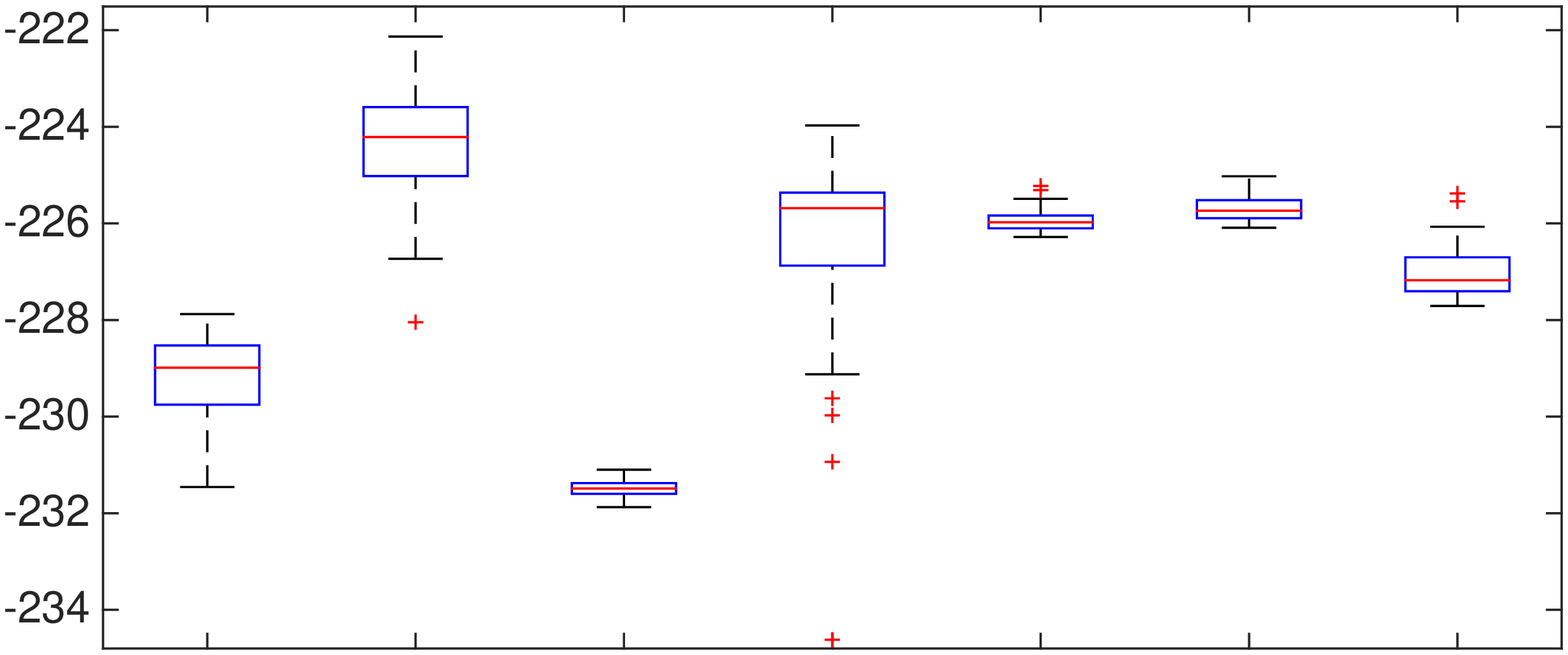}}
\put(8.5,6){\includegraphics[width=5cm, height=3cm]{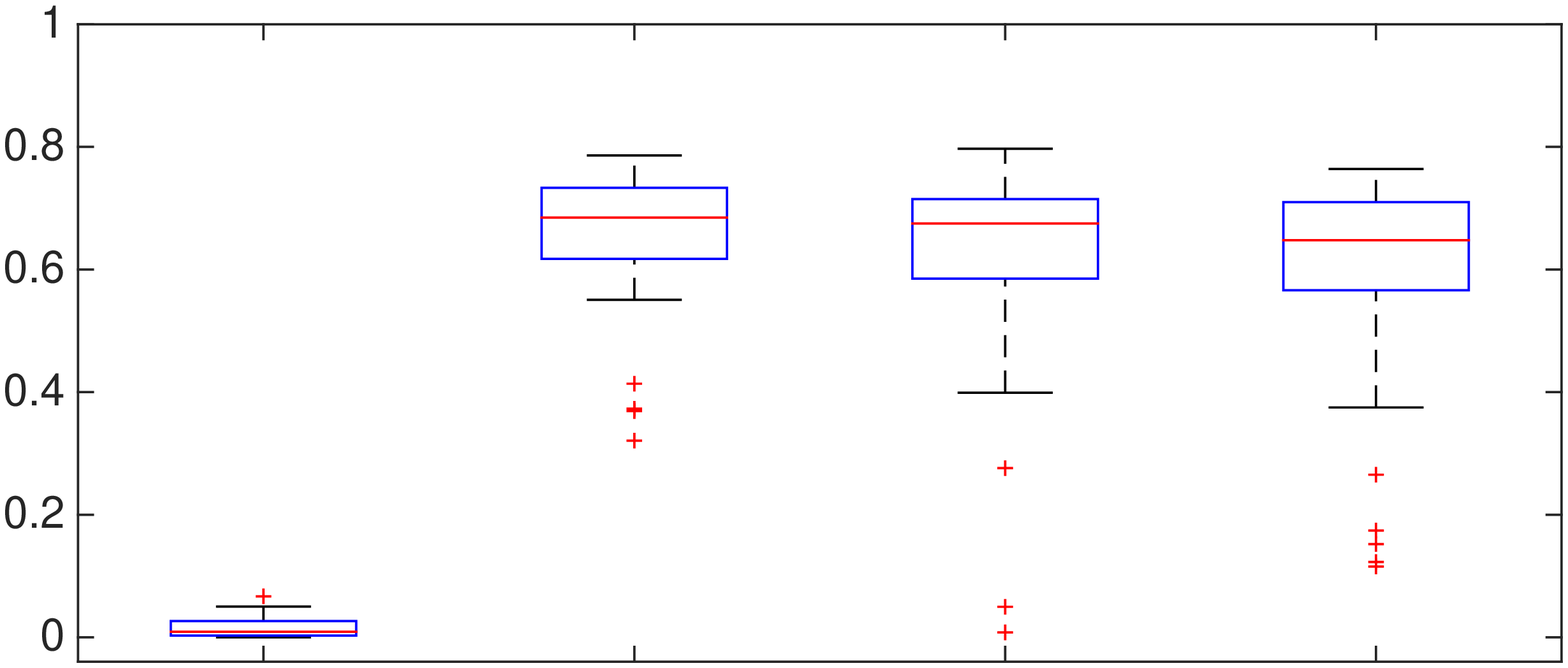}}
\put(8.5,3){\includegraphics[width=5cm, height=3cm]{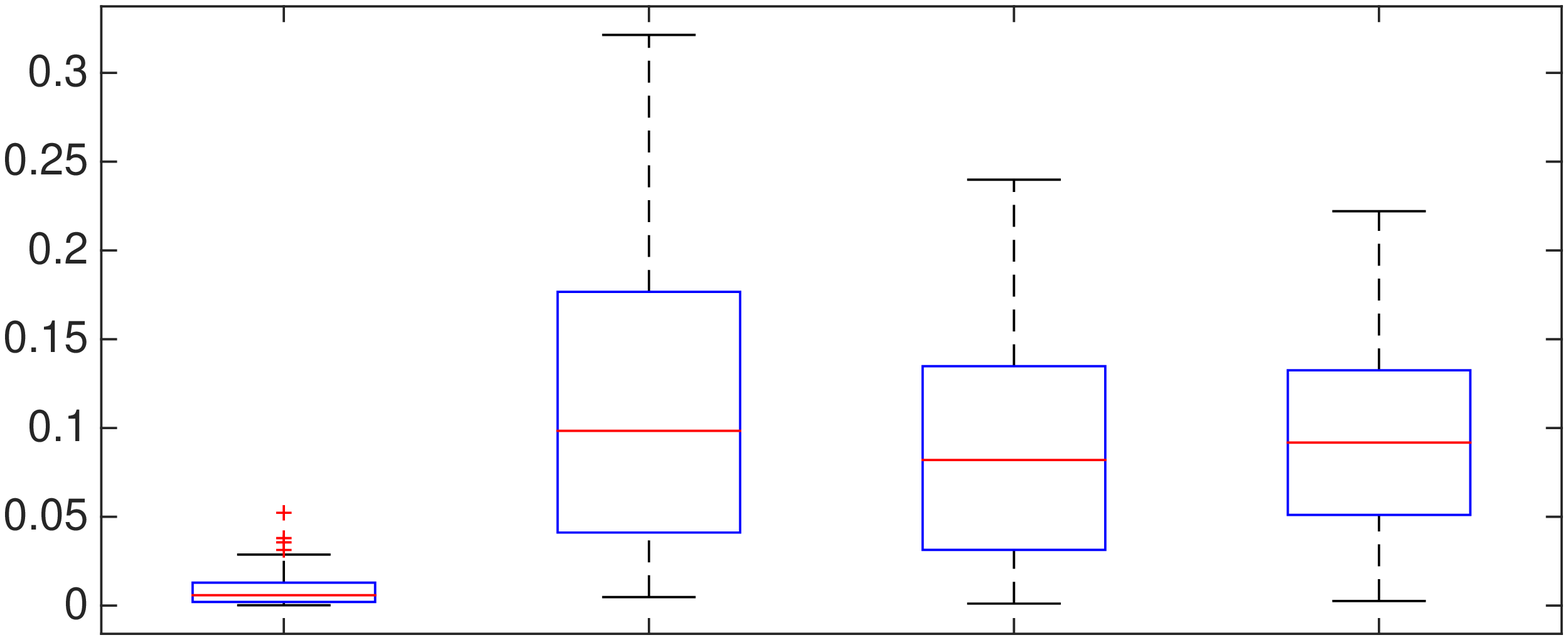}}
\put(8.5,0){\includegraphics[width=5cm, height=3cm]{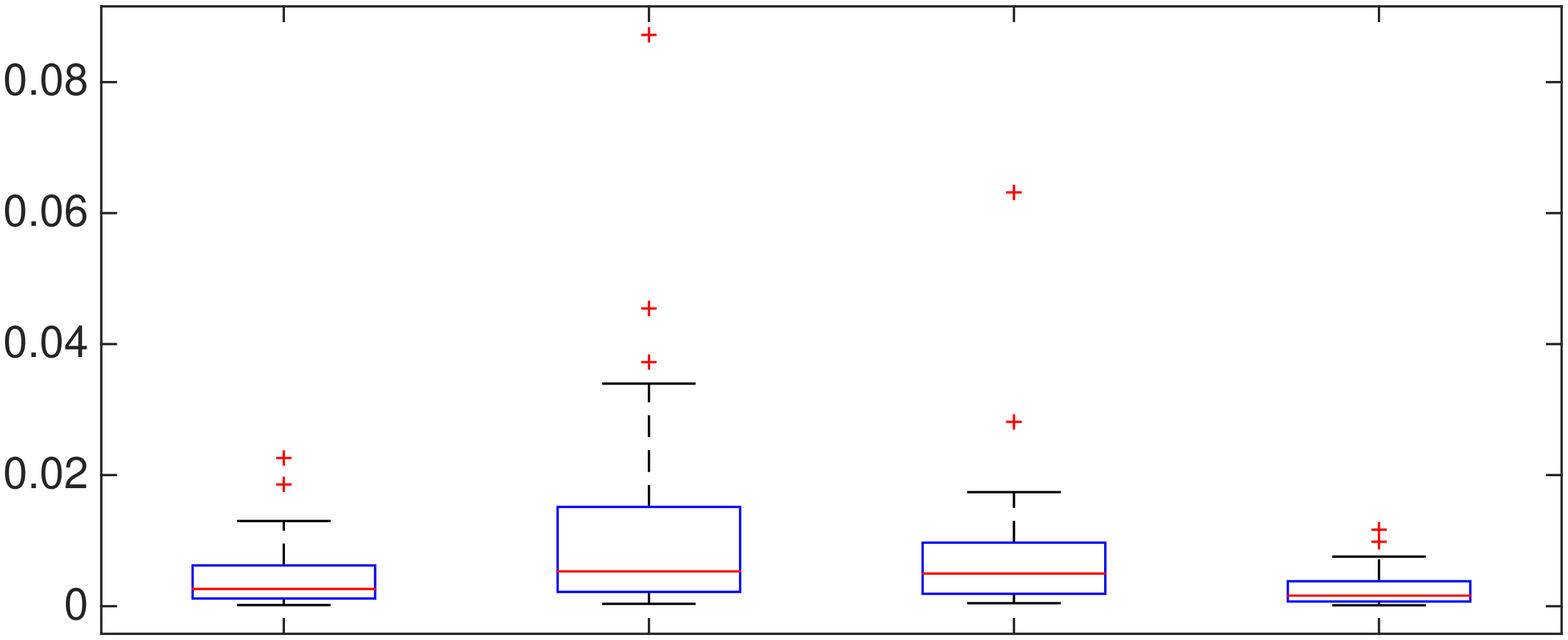}}

\rput(1.7,0.1){\footnotesize$\widehat{\mathfrak{E}}_{IS}$}  \rput(2.7,0.1){\footnotesize$\widehat{\mathfrak{E}}_{Ch}^*$} 
\rput(3.7,0.1){\footnotesize$\widehat{\mathfrak{E}}_{BS}$}  \rput(4.7,0.1){\footnotesize$\widehat{\mathfrak{E}}_{Ch}$}  
\rput(5.8,0.1){\footnotesize$\widehat{\mathfrak{E}}_{DS}$} \rput(6.7,0.1){\footnotesize$\widehat{\mathfrak{E}}^A_{DS}$} 
\rput(7.8,0.1){\footnotesize$\widehat{\mathfrak{E}}_{J_1}$}

\rput(1.7,3.1){\footnotesize$\widehat{\mathfrak{E}}_{IS}$}  \rput(2.7,3.1){\footnotesize$\widehat{\mathfrak{E}}_{Ch}^*$} 
\rput(3.7,3.1){\footnotesize$\widehat{\mathfrak{E}}_{BS}$}  \rput(4.7,3.1){\footnotesize$\widehat{\mathfrak{E}}_{Ch}$}  
\rput(5.8,3.1){\footnotesize$\widehat{\mathfrak{E}}_{DS}$} \rput(6.7,3.1){\footnotesize$\widehat{\mathfrak{E}}^A_{DS}$} 
\rput(7.8,3.1){\footnotesize$\widehat{\mathfrak{E}}_{J_1}$}

\rput(1.7,6.1){\footnotesize$\widehat{\mathfrak{E}}_{IS}$}  \rput(2.7,6.1){\footnotesize$\widehat{\mathfrak{E}}_{Ch}^*$} 
\rput(3.7,6.1){\footnotesize$\widehat{\mathfrak{E}}_{BS}$}  \rput(4.7,6.1){\footnotesize$\widehat{\mathfrak{E}}_{Ch}$}  
\rput(5.8,6.1){\footnotesize$\widehat{\mathfrak{E}}_{DS}$} \rput(6.7,6.1){\footnotesize$\widehat{\mathfrak{E}}^A_{DS}$} 
\rput(7.8,6.1){\footnotesize$\widehat{\mathfrak{E}}_{J_1}$}

\rput(9.7,0.1){\footnotesize $R_{IS}$} \rput(10.7,0.1){\footnotesize$R_{DS}$} 
\rput(11.7,0.1){\footnotesize$R^{A}_{DS}$} \rput(12.6,0.1){\footnotesize$R_{J_1}$}

\rput(9.7,3.1){\footnotesize $R_{IS}$} \rput(10.7,3.1){\footnotesize$R_{DS}$} 
\rput(11.7,3.1){\footnotesize$R^{A}_{DS}$} \rput(12.6,3.1){\footnotesize$R_{J_1}$}

\rput(9.7,6.1){\footnotesize $R_{IS}$} \rput(10.7,6.1){\footnotesize$R_{DS}$} 
\rput(11.7,6.1){\footnotesize$R^{A}_{DS}$} \rput(12.6,6.1){\footnotesize$R_{J_1}$}

\end{picture} \end{center}
\caption{
Boxplots of evidence estimates in log scale {\em (left, middle)} and effective sample sizes ratios {\em (right)}. Mixture models with {\em (top)} three, {\em (middle)} four, and {\em (bottom)} six Gaussian components are fitted to the galaxy dataset. One outlier of $\widehat{\mathfrak{E}}_{Ch}$ in the top-left panel is discarded. }  \label{Sim_galaxy} \end{figure}

The priors suggested by \citet{richardson:green:1997} are used for our simulation study:
\[ \begin{array} {r c  l}
\mu_1,\ldots,\mu_k &\sim& N(\bar{\bx},r^2/4) \\
\sigma_1^2,\ldots,\sigma_k^2 &\sim& IG(2,\beta) \\
\beta &\sim& G(0.2,10/r^2) \\
\lambda_1,\dots, \lambda_k &\sim& \text{Dirichlet}(1,\ldots , 1) 
\end{array} \]
where $\bar{\bx}$ and $r$ are the median and the range of $\bx$, respectively. Normal mixture models are fitted to both
datasets and estimates of $\log(\mathfrak{E}(k))$ and $R$ are summarized in Figures \ref{Sim_fishery} and
\ref{Sim_galaxy}. In general, a similar behaviour of $\log(\widehat{\mathfrak{E}}(k))$ and $R$ between the methods is
observed. For all cases, the dual importance sampling schemes ($\widehat{\mathfrak{E}}_{DS}$ and
$\widehat{\mathfrak{E}}_{DS}^A$) and $\widehat{\mathfrak{E}}_{J_1}$ agree with Chib's approach
($\widehat{\mathfrak{E}}_{Ch}$). Unless modes of the joint posterior distributions are clearly separated (e.g.,
$|\overline{\mathfrak{A}(k)}| \approx 1$), $\log(\widehat{\mathfrak{E}}^*_{Ch})$ is biased due to an improper
permutation correction. When a poor $q(\theta)$ is used for importance sampling, inaccurate approximations
result and the range of $\widehat{\mathfrak{E}}_{IS}$ estimates is much off from the other estimates.

\noindent Symptoms of the ``curse of dimensionality" can be observed. As $k$ increases, the effective sample size decreases
exponentially fast and the variation in the estimates increases. Given the complex shape of the posterior distribution,
the support common to $q(\theta)$ and $\pi^*_k(\theta|\bz)$ gets progressively smaller and $\widehat{\mathfrak{E}}_{BS}$
becomes less accurate, as shown in both figures. When $k=6$, the variation in the values of
$\widehat{\mathfrak{E}}_{Ch}$ is much larger than those of the estimate by dual importance sampling. When $J_1 \ll Jk!$,
$q$ does not provide a good approximation of the joint posterior and $\log(\mathfrak{E}_{J_1})$ is then biased. Due to a
fast increase of $k!$, fast increasing in CPU times is seen for all estimators in Table \ref{table_time_GalaxyFishery}. 

\noindent The reduction in the number of evaluated terms used to approximate $\widehat{\mathfrak{E}}(k)$ varies case by
case, as shown in Table \ref{table_GalaxyFishery}. When $k=4$ and $k=6$, components of the posterior distribution for
the galaxy data tend to have long flat tails and thus have higher chance to overlap each other. Consequently, the 
workload reduction is of lesser magnitude than for a model with a smaller number of components. Provided that some
functions are neglected for $\widehat{\mathfrak{E}}_{DS}^A$, there is some gain in computational efficiency as can be
seen in Table \ref{table_GalaxyFishery}.

\begin{table}[ht!]
\parbox{.45\linewidth}{
\centering
\begin{tabular} {c c | c c}
 $k$ & $k!$ & $|\mathfrak{A}(k)|$ &$\Delta(\mathfrak{A})$ \\ \hline
3 & 6 & 1.00 (0.00) & 0.25 (0.00) \\ 4 & 24 & 2.10 (0.76) & 0.18 (0.03) \\ \hline
\end{tabular} 
\\
(a) Fishery data
}
\hfill
\parbox{.45\linewidth}{
\centering
 \begin{tabular} {c c | c c }
 $k$ & $k!$ & $|\mathfrak{A}(k)|$ & $\Delta(\mathfrak{A})$ \\ \hline
  3 & 6 & 1.06 (0.24) & 0.26 (0.04) \\ 4 & 24 & 13.34 (5.35) &  0.60 (0.20) \\ 6 & 720 & 176.78 (75.31)& 0.32 (0.09) \\ \hline
\end{tabular} 
\\
(b) Galaxy data
}
\caption{Mean and standard deviation \emph{(values in brackets)} of approximate set sizes, $|\mathfrak{A}(k)|$, and the
reduction rate of a number of evaluated $h$-terms $\Delta(\mathfrak{A})$ as in (15) for (a) fishery and (b) galaxy
datasets. } \label{table_GalaxyFishery}
\end{table}

\begin{table}[ht!]
\[ \begin{array} {| c | c c | c c c |}
\hline 
\text{Estimator} & \multicolumn{2}{|c|}{\text{Fishery data}} & \multicolumn{3}{|c|}{\text{Galaxy data}} \\  
 & k=3 & k=4 & k=3 & k=4 & k=6 \\ \hline
\widehat{\mathfrak{E}}^*_{Ch} & 1.71 & 1.71 & 1.20 & 1.88 & 2.89 \\
\widehat{\mathfrak{E}}_{Ch} & 1.40 & 2.30 & 1.56 & 2.18 & 26.86 \\ 
\widehat{\mathfrak{E}}_{IS} & 12.23 & 14.60 & 13.47 & 14.83 & 48.74 \\
\widehat{\mathfrak{E}}_{DS} & 27.75 & 86.98 & 27.00 & 85.27 & 3.10\times 10^3 \\
\widehat{\mathfrak{E}}^{A}_{DS} & 18.14 & 30.45 & 18.28 & 52.49 & 1.33\times 10^3 \\
\widehat{\mathfrak{E}}_{J_1} & 28.19 & 90.11 & 26.75 & 87.19 & 244.10\\
\widehat{\mathfrak{E}}_{BS} & 4.92\times 10^3 & 6.71\times 10^3 & 4.21\times 10^3 & 3.14\times 10^3 & 7.32\times 10^3 \\ \hline
\end{array} \]
\caption{
Elapsed CPU time in seconds for evidences approximation of mixture models for fishery and galaxy datasets. } \label{table_time_GalaxyFishery}
\end{table}

\section{Discussion}

This paper considered evidence approximations by importance sampling
for mixture models and re-evaluated some of the known challenges resulting from high
multimodality in the posterior density. Importance sampling requires that the support of an importance function
encompasses the support of the posterior density to perform properly. In the specific case on mixture models, missing
some of the invariance under permutation function is likely to produce an unsuitable support hence, a poor estimate of
the evidence. 

\noindent In our study, exchangeable priors are used, which implies that
the posterior and marginal posterior densities exhibit
$k!$ symmetrical terms. Two marginal likelihood estimators are proposed here
and tested against other existing estimators. The first approach exploits the
permutation invariance of $\pi(\cdot|\bx,\bz^o)$ with a pointwise MLE, $\bz^o$,
to create an importance function. However, due to a poor resulting support,
this approach performs quite poorly in our simulation studies. Another poor
estimate is derived from Chib's method when the invariance by permutation is not reproduced in the sample
\citep{neal:2001}.

\noindent A second importance function is constructed by double  Rao--Blackwellisation, hence the denomination of
{\em dual importance sampling}. We demonstrate both methodologically and practically that this solution fits
the demands of mixture estimation. Moreover, introducing a suitable and implementable
approximation scheme, we show how to avoid the exponential increase in $k$ of the computational workload. The 
idea at the core of this approximation is to bypass negligible elements in the approximation thanks to the perfect symmetry of
the posterior density. When posterior modes are well-separated, the gain is of a larger magnitude than when those modes
strongly overlap.

\noindent Borrowing from the original approach in \citet{chib:1996}, dual importance sampling
can be extended to cases when conditional Gibbs sampling densities are not available in closed form. However, this
solution suffers from the curse of dimensionality, just like any other importance sampling estimator.

\noindent Alternative evidence approximation techniques could be considered for this problem, as exemplified in
\cite{fiel:wyse:2012}. For instance, {\em ensemble Monte Carlo}
samples from local ensembles that are extensions or compositions of the original, e.g., using
parallel tempering Monte Carlo methods. Extending this idea, Bayes factor approximations were proposed
using annealed importance sampling \citep{neal:2001} and power posteriors
\citep{fiel:pettitt:2008}. Further investigation is needed to characterize
the performances of those alternative solutions in the setting of mixture models and label switching.

\begin{acknowledgement}
We are most grateful to the Editorial Team of Bayesian Analysis for their
helpful suggestions and their support towards this revision. Financial support by CEREMADE, Universit\'e Paris-Dauphine
and Auckland University of Technology for a visit of Jeong Eun Lee is most appreciated. 
Christian Robert is
partially supported by the Agence Nationale de la Recherche (ANR, 212, rue de
Bercy 75012 Paris) through the 2010--2015 ANR-11-BS01-0010 grant
``Calibration'' and by a 2010--2015  Institut Universitaire de France senior chair.
\end{acknowledgement}

\end{document}